\documentclass[12pt,letterpaper]{JHEP3}

\usepackage{amscd,amsmath,amssymb,amsfonts,xspace,mathrsfs,amsthm}
\usepackage{color}
\let\normalcolor\relax
\usepackage{bbold}
\usepackage{latexsym}
\usepackage{graphicx}
\usepackage{dsfont}
\usepackage{color}
\usepackage{longtable}

  \newtheorem{conj}{Conjecture}

\numberwithin{equation}{section}

\def\det{\,{\rm det}\, }

\def\Im{\,{\rm Im}\,}

\def\({\left(}
\def\){\right)}
\def\[{\left[}
\def\]{\right]}
\def\<{\left\langle}
\def\>{\right\rangle}
\def\hf{{1\over 2}}
\def\haf{\textstyle{1\over 2}}

\newcommand{\I}{\mathrm{i}}

\newcommand{\cD}{\mathcal{D}}

\newcommand{\p}{\partial}

\newcommand{\cM}{\mathcal{M}}

\newcommand{\cN}{\mathcal{N}}

\newcommand{\cR}{\mathcal{R}}

\newcommand{\cJ}{\mathcal{J}}

\DeclareSymbolFont{AMSa}{U}{msa}{m}{n}
\DeclareSymbolFont{AMSb}{U}{msb}{m}{n}
\DeclareMathSymbol{\fieldR}{\mathalpha}{AMSb}{"52}

\newcommand{\cH}{\mathcal{H}}

\newcommand{\cA}{\mathcal{A}}

\newcommand{\nn}{\nonumber}

\newcommand{\eps}{\epsilon}

\newcommand{\IR}{\mathds{R}}

\newcommand{\IC}{\mathds{C}}
\newcommand{\IZ}{\mathds{Z}}

\newcommand{\IH}{\mathds{H}}

\newcommand{\IP}{\mathds{P}}

\newcommand{\Tr}{\mbox{Tr}}

\newcommand{\rank}{\mbox{rank}}

\newcommand{\q}{\mbox{q}}
\newcommand{\brq}{\bar\q}

\newcommand{\tT}{\tilde T}
\def\bea{\begin{eqnarray}}
\def\eea{\end{eqnarray}}
\def\be{\begin{equation}}
\def\ee{\end{equation}}
\def\ba{\begin{align}}
\def\ea{\end{align}}
\def\bse{\begin{subequations}}
\def\ese{\end{subequations}}

\def\ba{\bar a}

\def\btau{\bar \tau}

\def\hq{\hat q}

\def\tS{\tilde S}

\def\cij#1{c}
\def\ci#1{c}

\def\XXint#1#2#3{{\setbox0=\hbox{$#1{#2#3}{\int}$}
\vcenter{\hbox{$#2#3$}}\kern-.5\wd0}}

\def\Fcl{F^{\rm cl}}

\def\gamD#1{\tilde\gamma}

\def\CY{\mathfrak{Y}}

\DeclareMathOperator{\rk}{rk}

\def\cl0{\tilde c_0}

\def\bOm{\bar\Omega}

\def\whpsi{\widehat \psi}

\def\whh{\widehat h}

\def\cJr{\cJ^{\rm ref}}

\def\wr{w_{\rm ref}}
\def\mr{m_{\rm ref}}
\def\hr{h^{\rm ref}}

\def\whhr{\widehat h^{\rm ref}}

\def\Omi#1{\Omega^{(#1)}}
\def\bOmi#1{\bOm^{(#1)}}
\def\Zi#1{Z^{(#1)}}
\def\hi#1{h^{(#1)}}
\def\whhi#1{\whh^{(#1)}}

\def\cl{c^{(\ell)}}

\def\gama{\check\gamma}

\def\kk{K}
\def\dd{d}

\def\Bcm{B^{\rm cm}}

\title{Refinement and modularity of immortal dyons}

\author{Sergei Alexandrov$^\dagger$ and Suresh Nampuri$^\ddagger$
\\
$^\dagger${\it
Laboratoire Charles Coulomb (L2C), Universit\'e de Montpellier,
CNRS, F-34095, Montpellier, France}\\
\\
$^\ddagger${\it CAMGSD-IST, Av. Rovisco Pais 1, 1049-001 Lisboa,  Portugal }

\vspace*{2mm} {\tt e-mail:
\email{sergey.alexandrov@umontpellier.fr}, \email{nampuri@gmail.com}
}

\vspace*{-3mm}

}

\abstract{Extending recent results in ${\cal N}=2$ string compactifications, we propose that
the holomorphic anomaly equation satisfied by the modular completions of the generating functions of {\it refined} BPS indices
has a universal structure independent of the number ${\cal N}$ of supersymmetries.
We show that this equation allows to recover all known results about modularity
(under $SL(2,\mathds{Z})$ duality group) of BPS states in ${\cal N}=4$ string theory.
In particular, we reproduce the holomorphic anomaly characterizing the mock modular behavior of
quarter-BPS dyons and generalize it to the case of non-trivial torsion invariant.
}

\begin{document}

\section{Introduction}
\label{sec-intro}

The spectra of BPS states in theories with extended supersymmetry contain vital information about their
non-perturbative structures and strong coupling regimes. These in turn provide valuable insights into
the mathematical structures that encode the organization of the fundamental degrees of freedom in these theories.
Hence, determining these spectra constitutes an active avenue of research initiating a rich interplay
between theoretical and mathematical physics.
The BPS spectrum in any such theory is strongly constrained by symmetry requirements.
Specifically, in superstring theory compactifications the BPS spectrum is expected
to be invariant under the U-duality group acting
on charges carried by the BPS states.
The duality group also manifests itself in the same context, when it acts on
generating functions of BPS indices restricting them to be modular or mock modular forms.

The mock modular behavior \cite{Zwegers-thesis,MR2605321}
of the generating functions under $SL(2,\IZ)$ duality group has, in fact, been found to be a very generic phenomenon.
It features prominently  in the counting functions of black hole degeneracies in string compactifications
with $\cN=2$ and $\cN=4$ supersymmetry
\cite{Dabholkar:2012nd,Alexandrov:2016tnf,Alexandrov:2018lgp}, Vafa-Witten theory \cite{Vafa:1994tf,Dabholkar:2020fde},
Donaldson-Witten theory \cite{Korpas:2017qdo,Korpas:2019cwg}, moonshine phenomenon \cite{Cheng:2011ay} and numerous other setups.
Mock modularity implies that the holomorphic generating functions are not quite modular forms.
They have a very specific modular anomaly which
can be cancelled by the addition of a suitably chosen non-holomorphic term.
The resulting functions, called modular completions, do transform as modular forms,
but by construction fail to be holomorphic and satisfy certain holomorphic anomaly equations.
These modular completions and the associated anomaly equations are of paramount significance both from
a physical and mathematical standpoint.
On the one hand, they are building blocks of physical quantities such as partition functions
while on the other hand, they facilitate access to the precise modular transformation properties
of BPS indices which can notably be used to determine these indices exactly.

Recently, in \cite{Alexandrov:2018lgp} a quite general result about these modular completions
has been obtained in the context of D4-D2-D0 black holes in type IIA string theory
compactified on a generic Calabi-Yau (CY) threefold $\CY$.
This result took the form of a formula for expressing the modular completions $\whh_{p,\mu}(\tau,\btau)$
of the generating functions\footnote{The indices of $h_{p,\mu}$ label the divisor wrapped by D4-brane
$\cD=p^a\gamma_a\in\Lambda\equiv H_4(\CY,\IZ)$ and the residue flux $\mu\in \Lambda^\star/\Lambda$. }
$h_{p,\mu}(\tau)$ of BPS indices for these black holes,
evaluated at the large volume attractor point (see section \ref{subsec-D420} for precise definitions),
in terms of the generating functions $h_{p,\mu}(\tau)$ themselves.
This formula in particular implies that as soon as the divisor $\cD\subset \CY$ wrapped by D4-brane is reducible,
i.e. it can be decomposed into a sum with positive coefficients of other divisors,
the corresponding generating function is a vector valued (higher depth) mock modular form.
 This result was subsequently  generalized  in \cite{Alexandrov:2019rth} to include a refinement parameter $y=e^{2\pi \I z}$
conjugate to the angular momentum.
Although in the case of compact CY threefolds the refined BPS indices $\Omega(\gamma,y)$,
where $\gamma$ is an electromagnetic charge, are not protected by supersymmetry,
they can still be defined \cite{ks,Dimofte:2009bv,Manschot:2010qz}
and their generating functions possess very similar modular properties provided the parameter $z$
transforms as an elliptic parameter. As a result, the generating functions $\hr_{p,\mu}(\tau,z)$ of refined BPS indices
were shown to behave as {\it mock Jacobi forms} \cite{Dabholkar:2012nd}.

Although the formula for the modular completion has been found in the case of a {\it generic compact} CY,
i.e. a threefold with $SU(3)$ holonomy and not its proper subgroup, and for an {\it ample} divisor,
it turned out that it has a wider range of applicability. In particular, it can be extended to non-compact CYs
given by the canonical bundle over a projective surface $S$.
In this case it provides the modular completion of generating functions of (refined) Vafa-Witten invariants
of the surface $S$ with gauge group $U(N)$ \cite{Alexandrov:2019rth}.
In \cite{Alexandrov:2020bwg,Alexandrov:2020dyy} this result has been used to actually find these generating functions
as well as their completions explicitly, for {\it any} rank $N$ and $S=\IP^2$, Hirzebruch or del Pezzo.

In this paper we study another extension of that construction.
While it was originally formulated in the context of 4d $\cN=2$ supergravity, it is natural to ask whether it also captures
the modular properties of BPS indices in theories with larger amount of supersymmetry, which are obtained
by compactifying type II string theory either on $\CY=K3\times T^2$ or on $T^6$. This amounts to dropping the condition
that the holonomy group of $\CY$ is exactly $SU(3)$. Here we mostly restrict our analysis to the case $\CY=K3\times T^2$
corresponding to $\cN=4$ supersymmetry in four dimensions, and reserve the case $\CY=T^6$ for future work.

$\cN=4$ superstring compactifications have been extensively studied in the literature
and there is, at present, a substantial body of knowledge about their BPS indices
and the duality transformation properties of their generating functions.
As we review below in section \ref{sec-N4}, there are two types of BPS states for both of which the counting functions
are known exactly. For $\hf$-BPS states the generating function is a  weakly holomorphic $SL(2,\IZ)$ modular form of weight $-12$,
whereas the degeneracies of $\frac14$-BPS states are given by Fourier coefficients of $Z(\tau,z,\sigma)$,
the inverse of an Igusa cusp
Seigel modular form  of weight $10$, which transforms under the larger group $Sp(2,\IZ)$ \cite{Dijkgraaf:1996it}.
Expanding this function only in $\sigma$,
one finds that the coefficients $\psi_m(\tau,z)$ are Jacobi forms of weight $-10$ and index $m$
with respect to the usual modular group $SL(2,\IZ)$.
Although everything appears to be modular at this stage, the remarkable work \cite{Dabholkar:2012nd} revealed
a hidden mock modularity. It turns out that each meromorphic function $\psi_m$ has a canonical decomposition into
two parts,  $\psi_m^P$ and $\psi_m^F$. While the former only counts bound states,
the single centred black holes, known as {\it immortal dyons}, are encoded in the latter.
Both $\psi_m^P$ and $\psi_m^F$ are mock Jacobi forms and
the work \cite{Dabholkar:2012nd} provides the holomorphic anomaly equations satisfied by their completions, $\whpsi_m^P$ and $\whpsi_m^F$.

As will be shown below, all the results concerning the modular properties of BPS dyons under $SL(2,\IZ)$
can be derived from a simple extension of the holomorphic anomaly equation satisfied by the modular
completions $\whhr_{p,\mu}$ of the generating functions of refined BPS indices. This extension
allows to drop restrictions on $\CY$ and the divisor $\cD$.
The derivation is based on two observations. First, we note that the BPS indices counting black hole degeneracies
are given by helicity supertraces $B_{2K}$ \cite{Kiritsis:1997gu} with the center of mass contribution factored out.
Which helicity supertrace is relevant depends on the number of broken supersymmetries, but all of them can be obtained
from the refined BPS index $\Omega(\gamma,y)$ by taking appropriate number of derivatives with respect to $y$,
needed to kill fermionic zero modes, and then setting it to $y=1$.
In particular, in this way we arrive at generating functions which transform as (mock) modular forms of weight
\be
w=2\kk-3-\hf\rk,
\label{weight}
\ee
where $2\kk$ is the label of the relevant helicity supertrace and $\rk$ is the rank of the electric charge lattice $\Lambda_p$
assigned to the magnetic charge $p$.
Secondly, it is well known that in the chamber of the moduli space containing the large volume attractor point,
the indices counting BPS black holes with non-vanishing area
 in $\cN=4$ theory receive no contribution from bound states.
Combining these two observations, we get a direct way to relate the generating functions $\hr_{p,\mu}$ to
the functions $\psi_m^F$ or their $\hf$-BPS counterpart.

As a result, we show that the proposed holomorphic anomaly equation and the formula for weight \eqref{weight}
enable us to reproduce the generating function of $\hf$-BPS states.
For $\frac14$-BPS states we recover the correct modular weight as well and demonstrate
that the holomorphic anomaly equation satisfied by the functions $\whpsi_m^F$ counting immortal dyons
follows from the one satisfied by the completions $\whhr_{p,\mu}$.
Furthermore, whereas the holomorphic anomaly equation for $\whpsi_m^F$ derived in \cite{Dabholkar:2012nd}
holds only for charge vectors with the trivial torsion invariant $I(\gamma)=1$ (see \eqref{defI}),
we generalize it to arbitrary values of $I(\gamma)$. To the best of our knowledge,
this result, provided in eq. \eqref{mainanomaly}, did not appear in the literature before.

The organization of the paper is as follows. In the next section we recall  relevant aspects of
string compactifications with $\cN=4$ supersymmetry. In particular, we introduce the generating functions
counting immortal dyons and the holomorphic anomaly equation satisfied by their completions.
In section \ref{sec-N2} we review the construction of the completion of the generating functions of refined BPS indices
in $\cN=2$ string compactifications, extend it to cases with more supersymmetries,
and derive consequences of this extension in the unrefined limit.
Subsequently, in section \ref{sec-relation} we apply these results in the $\cN=4$ case where
we reproduce the counting function of $\hf$-BPS dyons,
derive the holomorphic anomaly equation for $\frac14$-BPS dyons,
demonstrate its consistency with \cite{Dabholkar:2012nd} and provide a generalization for $I(\gamma)>1$.
We conclude with a discussion in section \ref{sec-concl}.
The few appendices herein contain useful details on the chain of dualities between different formulations with $\cN=4$ supersymmetry,
modular transformations of generating functions, as well as restrictions on charges implied by the attractor mechanism.

\section{BPS states in $\cN=4$ string compactifications}
\label{sec-N4}

\subsection{Charges and generating functions}
\label{subsec-charge}

Type IIA string theory compactified on $K3\times T^2$ reduces in four dimensions to $\cN=4$ supergravity coupled to 22 vector multiplets.
The number 22 is the dimension of the second homology lattice of $K3$, which has signature (3,19)
and is composed by two lattices of $E_8$ and three two-dimensional even latices of signature (1,1), which will be denoted by $U_{1,1}$:
\be
H_2(K3,\IZ)=\Lambda_{3,19}=U_{1,1}^{\oplus 3}\oplus(-E_8)\oplus (-E_8).
\label{latK3}
\ee
The corresponding vector fields $A^i$, $i=1,\dots,22$,
arise from the ten-dimensional 3-form gauge field after its reduction on the 2-cycles of $K3$.
In addition, there are also 6 vector fields ($A^0,A^\flat,A^x$), $x=23,\dots 26$,
in the supergravity multiplet related to the geometry of the torus:
they originate from the ten-dimensional 1-form, the 3-form reduced on $T^2$,
the $B$-field reduced along the two circles of $T^2$ and the two KK gauge fields on $T^2$, respectively.
Thus, in total there are 28 gauge fields $A^I$, $I\in \{0,\flat,1,\dots,26\}$,
so that the electro-magnetic charge vector $\gamma=(p^I,q_I)$ has 56 components.

In fact, many features of the compactified theory, including its symmetries, are better seen in the dual heterotic formulation
which is obtained by compactifying heterotic superstring on $T^6$ \cite{Hull:1994ys}.
The 28 gauge fields now arise from the Cartan subalgebra of the rank 16 ten-dimensional gauge group,
the reduction of the B-field and
the KK gauge fields on $T^6$. They transform as a vector under T-duality group $SO(6,22)$,
whereas the full U-duality group
includes also $SL(2,\IZ)$ S-duality which acts on the heterotic axio-dilaton
\be
G_4(\IZ)=SL(2,\IZ)\times O(6,22;\IZ).
\ee
Note that in the type II formulation, the $SL(2,\IZ)$ factor is just the modular group of the torus $T^2$.

In this heterotic frame, the electro-magnetic charge belongs to the representation $(\bf{2},\bf{28})$
of the U-duality group and hence can be represented as a doublet of two vectors under the orthogonal group
\be
\gamma=\( \begin{array}{c} Q_I \\ P_I \end{array}\)\in \Lambda_{em}=\Lambda_e\oplus \Lambda_m,
\label{charge-het}
\ee
where
\be
\Lambda_e=\Lambda_{6,22}=U_{1,1}^{\oplus 6}\oplus(-E_8)\oplus (-E_8)
\label{Lambdae}
\ee
and $\Lambda_m=\Lambda_e^* =\Lambda_e$ because $\Lambda_e$ is unimodular.
The bilinear form on $\Lambda_e$ is given by an $O(6,22)$ invariant matrix
\be
\eta_{IJ}=\(\begin{array}{cc}
I_{1,1}^{\oplus 6} & 0
\\
0 & -C_{16}
\end{array}\),
\qquad
I_{1,1}=\(\begin{array}{cc}
0 & 1
\\
1 & 0
\end{array}\),
\label{matL}
\ee
where $C_{16}$ is the Cartan matrix of $E_8\times E_8$. Using this bilinear form,
one can define three T-duality invariants and one their U-duality invariant combination
\be
\begin{split}
n=&\, \hf \, Q^2=\hf\, Q_I \eta^{IJ} Q_J,
\\
m=&\, \hf \, P^2=\hf\, P_I \eta^{IJ} P_J,
\qquad \qquad
\Delta=4mn-\ell^2.
\\
\ell=&\, Q\cdot P= Q_I \eta^{IJ} P_J,
\end{split}
\label{invar}
\ee
Besides, there is also another U-duality invariant, the torsion, which is defined as follows \cite{Dabholkar:2007vk}
\be
I(\gamma)=\gcd\{Q_IP_J-Q_J P_I\}.
\label{defI}
\ee
Taken together, these invariants are sufficient to characterize uniquely the duality orbit of the charge vector
\cite{Banerjee:2007sr,Banerjee:2008ri}.

The heterotic frame is particularly convenient to classify BPS states.
There are two classes of them: $\hf$-BPS and $\frac14$-BPS.
The former are characterized by the charges \eqref{charge-het}
which have parallel electric and magnetic charge vectors, $Q\parallel P$.
This implies that there is a duality frame where one of them, e.g. $P$, can be set to zero.
Thus, the $\hf$-BPS index, which is invariant under the full U-duality group,
depends only on one T-duality invariant $n$ and, moreover,
it is known to be independent of the moduli.
Thus, the $\hf$-BPS spectrum can be encoded into a generating function of one variable
\be
\Zi{4|2}(\tau)=\sum_{n=-1}^\infty \Omi{4|2}(n)\, \q^n,
\label{countd}
\ee
where $\q=e^{2\pi\I \tau}$ and we accepted the convention, which will be extensively used below,
that the upper index of type $(\cN|r)$
refers to $\frac1r$-BPS states in a theory with $\cN$ extended supersymmetry.
The generating function \eqref{countd} has been found explicitly \cite{Dabholkar:1989jt}
and is given by a modular form of weight $-12$,
\be
\Zi{4|2}(\tau)=q^{-1}\prod_{n=1}^\infty (1-\q^n)^{-24}=\eta(\tau)^{-24},
\label{Z12eta}
\ee
where $\eta(\tau)$ is the Dedekind eta function.
The coefficients in \eqref{countd} can be computed in terms of the partition function
\be
\Omi{4|2}(n)=p_{24}(n+1),
\ee
where $p_{24}(N)$ is the number of partitions of a positive integer $N$ into 24 colored integers.

The spectrum of $\frac14$-BPS states, which have $Q$ and $P$ non-parallel, is much more complicated.
First of all, their degeneracies depend on all duality invariants introduced above
so that we can write\footnote{The index $I$ here denotes the value of the discrete torsion invariant
and should not be confused with the index labelling charges.}
$\Omi{4|4}(\gamma)=\Omi{4|4}_I(n,m,\ell)$.
Thus, they can be combined into a generating function of three variables labelled by the torsion invariant\footnote{We introduced
the sign $(-1)^\ell$ consistently with \cite{Shih:2005uc}
where it was argued on the basis of a connection between 5d and 4d black holes \cite{Gaiotto:2005gf}.
We will see below that it is also crucial for the relation with the generating functions of refined BPS indices.}
\be
\Zi{4|4}_I(\tau,z,\sigma)=\sum_{n,m,\ell}(-1)^\ell\, \Omi{4|4}_I(n,m,\ell)\, \q^n \, y^\ell\,  p^m,
\ee
where $y=e^{2\pi\I z}$, $p=e^{2\pi\I\sigma}$ and $\q$ is as in \eqref{countd}.
The famous result of \cite{Dijkgraaf:1996it} expresses this function for $I=1$ in terms of the so-called Igusa cusp form
\be
\Zi{4|4}_1(\tau,z,\sigma)=\frac{1}{\Phi_{10}(\tau,z,\sigma)}\, ,
\label{Igusa}
\ee
which is a modular form with respect to $Sp(2,\IZ)$ acting on the Seigel upper half-plane parametrized by $\tau$, $z$ and $\sigma$.
We will not use this fact in this paper since our analysis is restricted
to the modularity with respect to the usual S-duality group $SL(2,\IZ)$.
The degeneracies of $\frac14$-BPS states with $I>1$ have been found in \cite{Dabholkar:2008zy}
and can be expressed through those with $I=1$ determined by \eqref{Igusa},
\be
\Omi{4|4}_I(n,m,\ell)=\sum_{\dd | I} \dd \, \Omi{4|4}_1\(n,\frac{m}{\dd^2},\frac{\ell}{\dd}\).
\label{Om14-I}
\ee

Another complication compared to the $\hf$-BPS case is that the indices $\Omi{4|4}_I$ are actually moduli dependent.
The physical reason for this is that a $\frac14$-BPS state can be a bound state of two $\hf$-BPS ones \cite{Dabholkar:2009dq}.
Such bound states do not exist throughout all the moduli space and decay after crossing the lines of their marginal stability \cite{Sen:2007vb}.
Mathematically, this dependence is manifested by the existence of second order poles in the generating function \eqref{Igusa}.
As a result, its Fourier coefficients are not defined uniquely, but depend on the integration contour
in the Seigel upper half-plane which in turn is determined by the moduli \cite{Cheng:2007ch}.

While the heterotic frame is very convenient for formulating results about BPS states, for our purposes we need
to relate it to the type IIA frame. To this end, we express the charge vector \eqref{charge-het} through the charges $(p^I, q_I)$
introduced in the beginning of this section.
The resulting charge vector reads \cite{Dabholkar:2005dt}
\be
\gamma=\(\begin{array}{ccc}
q_0, & -p^\flat, & q_\alpha
\\
q_\flat, & p^0, & \eta_{\alpha\beta}p^\beta
\end{array}\),
\label{chD420}
\ee
where $\alpha=1,\dots,26$. In appendix \ref{ap-dual} we recall the chain of dualities leading to this result.

\subsection{Mock modularity of immortal dyons}
\label{subsec-mockimmort}

Let us return to $\frac14$-BPS states and expand their generating function \eqref{Igusa}
for the trivial torsion invariant in Fourier series in $\sigma$
\be
\frac{1}{\Phi_{10}(\tau,z,\sigma)}=\sum_{m=-1}^\infty\psi_m (\tau,z)\, p^m.
\label{expZsigma}
\ee
From the modularity of $\Phi_{10}$ with respect to $Sp(2,\IZ)$ it follows that the coefficients $\psi_m (\tau,z)$ are
Jacobi forms of weight $-10$ and index $m$ (see \eqref{Jacobi} for the definition of a general Jacobi form).
Note that if we considered the generating function $\Zi{4|4}_I$ with $I>1$ and performed a similar expansion,
due to \eqref{Om14-I} the corresponding coefficients would be Jacobi forms only with respect to
a congruence subgroup $\Gamma^0(I)\subset SL(2,\IZ)$  \cite{Dabholkar:2008zy} defined by
\be
\Gamma^0(I)=\left\{ \(\begin{array}{cc} a & b \\ c & d\end{array}\)\in SL(2,\IZ)\ : \ b=0\mod I\right\}.
\ee

The Jacobi forms $\psi_m$ inherit the double poles from $\Zi{4|4}_1$ and hence are meromorphic functions in $z$.
In \cite{Dabholkar:2012nd} it was shown that they admit a canonical decomposition
\be
\psi_m=\psi^P_m+\psi^F_m,
\label{decomppsi}
\ee
where $\psi^P_m$ contains the ``polar" part of the original function and is captured by the so-called
Appell-Lerch sum, whereas $\psi^F_m$ is holomorphic.
This decomposition was also given a physical interpretation: all contributions of
single centered black holes are encoded in $\psi^F_m$ while $\psi_M^P$ only contains contributions of bound states.
Since the bound states decay at lines of marginal stability,
it is possible to choose a region of moduli space where $\psi^P_m$ vanishes.
Single-centered black holes exist in all regions of moduli space and hence,
$\psi^F_m$ can be viewed as the counting function of immortal dyons.

An important result of \cite{Dabholkar:2012nd} is that both functions  $\psi^P_m$ and $\psi^F_m$ are mock Jacobi forms,
i.e. their modular transformations are anomalous.
The corresponding modular anomalies must be equal and opposite so as to cancel
each other on adding the corresponding mock Jacobi forms to give a fully modular object as in \eqref{decomppsi}.
Furthermore, there is a canonical way to construct modular completions
$\whpsi^P_m$ and $\whpsi^F_m$ which do transform as Jacobi forms, but are not holomorphic.
In particular, the modular completion $\whpsi^F_m$ was shown to satisfy the following
holomorphic anomaly equation\footnote{In \cite[Eq.(1.7)]{Dabholkar:2012nd} this equation was given with a different coefficient:
the denominator $8\pi\I$ appeared under square root. However, it must follow from applying Eq. (7.4) with $k=3/2$
to the function given in Eq. (9.5) in that paper. Collecting all coefficients, one does find $1/(8\pi\I)$, as we wrote in \eqref{holanom}.}
\be
\tau_2^{3/2}\p_{\btau}\whpsi^F_m(\tau,z)=\frac{\sqrt{m}}{8\pi\I}\, \frac{\Omi{4|2}(m)}{\eta(\tau)^{24}}
\sum_{\ell=0}^{2m-1}\overline{\theta_{m,\ell}(\tau,0)}\, \theta_{m,\ell}(\tau,z)\equiv \cA_m(\tau,z),
\label{holanom}
\ee
where
\be
\theta_{m,\ell}(\tau,z)=\sum_{r\in 2m\IZ +\ell}\q^{\frac{r^2}{4m}}\, y^r.
\label{deftheta}
\ee
This is the result which will be reproduced below, and generalized to include a non-trivial torsion,
from a completely different approach starting from
the generating functions of D4-D2-D0 black holes degeneracies in $\cN=2$ compactifications.

\section{Refined BPS index and modularity}
\label{sec-N2}

\subsection{D4-D2-D0 black holes in $\cN=2$ setup}
\label{subsec-D420}

Let us consider type IIA string theory compactified on a generic compact CY threefold $\CY$ with $SU(3)$ holonomy group.
At low energies it is described by $\cN=2$ supergravity coupled
to $h_{1,1}(\CY)=b_2(\CY)$ vector multiplets and $h_{2,1}(\CY)+1$ hypermultiplets.
The $\hf$-BPS states in this theory are labelled by electro-magnetic charge $\gamma=(p^0,p^a,q_a,q_0)$ with $a=1,\dots,b_2$
whose entries correspond to D6, D4, D2 and D0-brane charges, respectively, and at large coupling are realized as
supersymmetric black holes.
They are counted (with sign) by BPS indices $\Omi{2|2}(\gamma)$ which are known to coincide with
the generalized Donaldson-Thomas (DT) invariants of the CY threefold.
In fact, below we will deal mostly with their refined version
\be
\Omega(\gamma,y)=\Tr_{\cH'_\gamma}(-y)^{2J_3},
\label{defOmr}
\ee
which reduces to $\Omi{2|2}(\gamma)$ at $y=1$.
Here $J_3$ is a Cartan generator in the massive little group in 3+1 dimensions, $\cH'_\gamma$ is the Hilbert space
of states graded by the charge $\gamma$ with the center of mass degrees of freedom excluded, and $y$ is a refinement parameter.
In particular, we will be interested in the modular properties of the generating functions of the refined indices.

While in string theory on $K3\times T^2$ the modular symmetry can be trivially
identified with large diffeomorphisms of the torus, one may wonder where it comes from for type IIA on a generic CY.
It can be revealed by compactifying this theory on a circle.
Indeed, the resulting theory has two dual formulations:
it can be viewed either as M-theory on $\CY\times T^2$ or, after applying T-duality along the circle,
as type IIB string theory on $\CY\times S^1$. In both formulations the modular symmetry is evident\footnote{We note
that the $SL(2,\IZ)$ modular group revealed by this construction is universal to all type II compactifications
and it is {\it distinct} from the electric-magnetic duality group in the heterotic frame
which appears in the case of $\cN =4$ compactifications.}
and must be realized as an isometry of the moduli space $\cM_{3d}$ in three dimensions \cite{Alexandrov:2008gh}.
Given that the BPS states in four dimensions induce instanton corrections weighted by $\Omi{2|2}(\gamma)$
to the metric on $\cM_{3d}$,
this modular isometry constraint imposes non-trivial restrictions on the BPS indices.

To proceed further we need to restrict to the vanishing D6-brane charge $p^0$.
The reason is that $SL(2,\IZ)$ mixes D6-branes wrapped on $\CY\times S^1$ with KK-instantons (or in the T-dual picture,
D5 and NS5-branes both wrapped on $\CY$). Whereas the exact description of D-instantons
is well-known by now \cite{Alexandrov:2008gh,Alexandrov:2009zh,Alexandrov:2011va}, the understanding of NS5-brane instantons
remains incomplete (see \cite{Alexandrov:2010ca,Alexandrov:2014mfa,Alexandrov:2014rca}),
which makes it difficult to draw precise conclusions for the modular properties of BPS indices with non-vanishing $p^0$.

Thus, we consider BPS states with the charge $\gamma=(0,p^a,q_a,q_0)$ satisfying
the following quantization conditions \cite{Alexandrov:2010ca}:
\be
\label{fractionalshiftsD5}
p^a\in\IZ ,
\qquad
q_a \in \IZ  + \frac12 \,\kappa_{abc}p^b p^c  ,
\qquad
q_0\in \IZ-\frac{1}{24}\,c_{2,a} p^a.
\ee
Here $c_{2,a}$ are components of the second Chern class of $\CY$ and $\kappa_{abc}$ are the intersection numbers on $\Lambda=H_4(\CY,\IZ)$.
Furthermore, the divisor $\cD=p^a\gamma_a$, where $\gamma_a$ is a basis of $\Lambda$, wrapped by D4-brane must be ample, i.e.
belong to the K\"ahler cone defined, in terms of a convenient notation $(lkp)=\kappa_{abc}l^a k^b p^c$, by
\be
\label{khcone}
p^3> 0,
\qquad
(r p^2)> 0,
\qquad
k_a p^a > 0,
\ee
for all divisor classes $r^a \gamma_a \in H_4(\CY,\IZ)$ with $r^a>0$, and
curve classes $k_a \gamma^a \in H_2(\CY,\IZ)$ with $k_a>0$.
Under these conditions, the charge $p^a$ induces a quadratic form $\kappa_{ab}=\kappa_{abc} p^c$ of signature $(1,b_2-1)$
on the lattice $\Lambda$.
This quadratic form allows to embed $\Lambda$ into $\Lambda^*=H_2(\CY,\IZ)$, but the map $\epsilon^a \mapsto \kappa_{ab} \epsilon^b$
is in general not surjective, the quotient $\Lambda^*/\Lambda$ being a finite group of order $|\det\kappa_{ab}|$.

As in $\cN=4$ string theory, the BPS indices carry dependence on CY moduli $z^a$ due to the existence of bound states and
the wall-crossing phenomenon \cite{Denef:2000nb}. But while in $\cN=4$ there are only bound states with two constituents
\cite{Dabholkar:2009dq}, in the $\cN=2$ setup the number of constituents can be arbitrary.
Given that the modular group acts on $z^a$ and thus can map from one chamber of the moduli space $\cM$ to another,
the complicated wall-crossing pattern makes it difficult to expect any simple modular properties from the BPS indices
considered at arbitrary point in $\cM$.

This problem is solved by considering the BPS indices evaluated at
the large volume attractor point $z^a=z_\infty^a(\gamma)$,
\be
z_\infty^a(\gamma)=\mathop{\rm lim}\limits_{\lambda\to\infty}(-\kappa^{ab}q_b+\I\lambda p^a).
\label{zatr}
\ee
Such indices $\Omi{2|2}_\star(\gamma)\equiv \Omi{2|2}(\gamma;z^a_\infty(\gamma))$
as well as their refined version possess a set of important properties:
\begin{itemize}
\item
{\bf Spectral flow symmetry.} They are invariant under spectral flow transformations
acting on the D2 and D0 charges via
\be
\label{flow}
q_a \mapsto q_a - \kappa_{ab}\epsilon^b,
\qquad
q_0 \mapsto q_0 - \epsilon^a q_a + \frac12\, \kappa_{ab}\epsilon^a \epsilon^b
\ee
with $\epsilon^a\in\Lambda$. These transformations leave invariant the combination
\be
\hq_0=q_0-\hf\, \kappa^{ab} q_a q_b,
\label{defhq0}
\ee
where $\kappa^{ab}$ is the inverse of $\kappa_{ab}$, and allow a decomposition of the D2-brane charge
into the spectral flow parameter $\epsilon$ and the residue class $\mu\in \Lambda^*/\Lambda$
\be
q_a=\kappa_{ab}\epsilon^b +\mu_a +\hf\, \kappa_{ab}p^b,
\qquad
\eps^a\in \IZ.
\label{sfdecomp}
\ee
As a result, due to the spectral flow symmetry, the BPS indices evaluated at \eqref{zatr}
depend only on $\hat q_{0}$, $p^a$ and $\mu_a$ so that we can write (in the presence of refinement)
\be
\Omega_\star(\gamma,y)=\Omega_{p,\mu}(\hq_0,y).
\ee

\item
{\bf Bogomolov bound.} The invariant charge $\hq_0$ \eqref{defhq0} is bounded from above because
DT invariants are known to vanish for
$\hat q_0 >\hat q_0^{\rm max}=\frac{1}{24}\chi(\cD)=\tfrac{1}{24}(p^3+c_{2,a}p^a)$.

\item
{\bf Relation to single-centered black holes.}
The index $\Omega_{p,\mu}(\hq_0,y)$ is known to be closely related to the one counting single-centered
black holes. Although they do not coincide  \cite{Alexandrov:2018iao},
the difference between these two indices is only due to the so-called scaling solutions,
i.e. multi-centered black holes whose constituents can become arbitrarily close to
each other and are allowed to exist in the attractor chamber.
An important feature of these scaling solutions is that they comprise of {\it at least} 3 centers
\cite{Denef:2007vg,Bena:2012hf}.

\end{itemize}
These properties allow to define a natural generating function
\be
\hr_{p,\mu}(\tau,z) =\sum_{\hat q_0 \leq \hat q_0^{\rm max}}
\frac{\bOm_{p,\mu}(\hat q_0,y)}{y-y^{-1}}\,\q^{-\hat q_0  },
\label{defhDTr}
\ee
where we represented the refinement parameter as $y=e^{2\pi\I z}$ and
introduced rational invariants which differ from the integer valued ones only for non-primitive charges,
\be
\label{defbOm}
\bOm(\gamma,y) = \sum_{d|\gamma} \frac{y-y^{-1}}{d(y^d-y^{-d})}\, \Omega(\gamma/d,y^d) .
\ee
In the refined case, the rational invariants appeared first in \cite{Manschot:2010qz} where they were shown to
have simplified wall-crossing properties. It is crucial to use them in the generating function for the latter
to possess nice modular properties, and we will see that they are also important for agreement with the results presented in
the previous section.

In \cite{Alexandrov:2019rth}, on the basis of the unrefined construction resulting from the analysis of
the three-dimensional moduli space $\cM_{3d}$ \cite{Alexandrov:2018lgp}, it was argued that
that the generating functions \eqref{defhDTr} of refined BPS indices transform under modular transformations
\be
\tau\to \frac{a\tau+b}{c\tau+d}\, ,
\qquad
z\to \frac{z}{c\tau+d}\, ,
\qquad
\(\begin{array}{cc}
a & b \\ c & d
\end{array}\)\in SL(2,\IZ)
\label{transf-tz}
\ee
as vector valued (higher depth) mock Jacobi forms\footnote{More precisely, if the divisor $\cD$ is a sum of $n$ {\it irreducible}
divisors, then $\hr_{p,\mu}$ is a vector valued mock Jacobi form of depth $n-1$. In particular, this means that for $\cD$ irreducible
the generating function is a Jacobi form and for $n=2$ it is the standard mock Jacobi form.}
of weight $\wr=-\hf b_2$ and index $\mr(p)=-\(\frac{1}{6} p^3 + \frac{1}{12} c_{2,a}p^a\)$.
Their precise transformations can be found in appendix \ref{ap-modular}.
This result implies that there is canonical way to construct a non-holomorphic modular completion $\whhr_{p,\mu}$
which transforms as a true Jacobi form of the same weight and index,
and the main result of \cite{Alexandrov:2019rth} was a formula for such completion expressing it as a series in $\hr_{p_i,\mu_i}$
with $\sum_i p_i^a=p^a$.
We will be mainly interested in the holomorphic anomaly equation satisfied by $\whhr_{p,\mu}$ which was also derived
in \cite{Alexandrov:2019rth} and takes the following form
\be
\p_{\bar\tau}\whhr_{p,\mu}(\tau,z)= \sum_{n=2}^\infty
\sum_{\sum_{i=1}^n \gama_i=\gama}
\cJr_n(\{\gama_i\},\tau_2,y)
\, e^{\pi\I \tau Q_n(\{\gama_i\})}
\prod_{i=1}^n \whhr_{p_i,\mu_i}(\tau,z),
\label{exp-derwh}
\ee
where $\gama=(p^a,q_a=\mu_a+\hf\, \kappa_{ab}p^b)$, $\gama_i=(p_i^a,q_{i,a})$, the electric charges $q_{i,a}$ are decomposed as in
\eqref{sfdecomp} with the quadratic form $\kappa_{i,ab}=\kappa_{abc}p_i^c$, and
\be
Q_n(\{\gama_i\})= \kappa^{ab}q_a q_b-\sum_{i=1}^n\kappa_i^{ab}q_{i,a} q_{i,b} \, .
\label{defQlr}
\ee
Each term in \eqref{exp-derwh} has the form of the product of the completions corresponding to charges $p_i^a$,
multiplied by a theta series determined by the quadratic form $Q_n$ and the kernel $\cJr_n$.
It is important that due to the restrictions on the divisors $\cD_i=p_i^a\gamma_a$, all $\kappa_{i,ab}$ are of Lorentzian signature
so that $Q_n$ has rank $(n-1)b_2$ and signature $((n-1)(b_2-1),n-1)$.

Of course, all non-trivialities are hidden in the kernels $\cJr_n$. They are expressed through the so called generalized
error functions \cite{Alexandrov:2016enp,Nazaroglu:2016lmr}, but we will not need their explicit form in this work
and refer an interested reader to \cite{Alexandrov:2019rth}.
The only fact which is relevant for us is that they have a zero of order $n-1$ at $y=1$,
i.e.
\be
\mathop{\rm lim}\limits_{y\to 1}\Bigl[(y-y^{-1})^{1-n} \cJr_n(\{\gama_i\},\tau_2,y)\Bigr]=\cJ_n(\{\gama_i\},\tau_2)
\label{unreflim-cJ}
\ee
is finite and well defined. In particular, one can show that
\be
\cJ_2(\{\gama_1,\gama_2\},\tau_2)=\frac{(-1)^{\gamma_{12}}\sqrt{(pp_1p_2)}}{8\pi\I (2\tau_2)^{3/2}}
\, e^{-\frac{2\pi\tau_2 \gamma_{12}^2}{(pp_1p_2)}},
\label{cJ2}
\ee
where
\be
\gamma_{12}=q_{1,a}p_2^a -q_{2,a}p_1^a
\ee
is the Dirac product on the charge lattice (for vanishing D6-brane charge).
Taking into account the presence of the factor $(y-y^{-1})^{-1}$ in the definition of
the generating functions \eqref{defhDTr},
the property \eqref{unreflim-cJ} ensures that the holomorphic anomaly equation \eqref{exp-derwh} has a well defined
unrefined limit $y\to 1$.
Provided the generating functions of {\it unrefined} BPS indices are defined by
\be
h_{p,\mu}(\tau)=\mathop{\rm lim}\limits_{z\to 0}\Bigl[(y-y^{-1})\hr_{p,\mu}(\tau,z)\Bigr]
=\sum_{\hat q_0 \leq \hat q_0^{\rm max}}
\bOmi{2|2}_{p,\mu}(\hat q_0)\,\q^{-\hat q_0  },
\label{defhDT}
\ee
their completions $\whh_{p,\mu}$ satisfy a holomorphic anomaly equation similar to \eqref{exp-derwh}
with $\whhr_{p,\mu}$ and $\cJr_n$ replaced by $\whh_{p,\mu}$ and $\cJ_n$, respectively.

\bigskip

Our goal is to extend the results just described to more general compactifications which include in particular
CY threefolds with a reduced holonomy group like $K3\times T^2$.
One immediately notices some similarities in the (mock) modular behavior of the generating functions
defined in this and the previous sections. However, a closer inspection reveals important differences:
\begin{itemize}
\item
While the anomaly equations \eqref{holanom} and \eqref{exp-derwh} both imply mock modularity of the holomorphic generating functions,
the r.h.s. of the former is only {\it quadratic} in BPS indices, whereas in the latter one sums over
all possible decompositions of the charge into any number of constituents with the only condition of $\cD_i$ being ample.

\item
In the case of $\CY=K3\times T^2$, the electric charge lattice is larger than the second homology lattice $H_2(\CY,\IZ)$
and its signature is {\it not} Lorentzian.

\end{itemize}
In the next subsections we show how these differences can be incorporated into a general framework
allowing to establish a precise correspondence with the well known results in $\cN=4$ string compactifications.

\subsection{Helicity supertraces and refined index}

The differences delineated above between the $\mathcal{N}=2$ and $\mathcal{N}=4$ cases
are not too surprising in light of the fact that we are trying to compare theories with different amount of supersymmetry.
Furthermore, it is important to note that the BPS indices in the two theories are genuinely different quantities,
irrespective of whether they count BPS states with the same or different number of conserved supersymmetries.
These indices, similar to the Witten index, perform a weighted count (with signs)
of the short and intermediate multiplets in the corresponding supersymmetric theories and are precisely defined
in terms of {\it helicity supertraces} \cite{Kiritsis:1997gu},
\be
B_{2\kk}(\cR)=\Tr_\cR\Bigl[ (-1)^{2J_3} J_3^{2\kk}\Bigr],
\label{defHS}
\ee
where $\cR$ is a representation of the supersymmetry algebra.
The insertion of each power of $J_3$ in the trace soaks up $2$ fermionic zero modes.
Each fermionic zero mode corresponds to a broken supercharge and all of them should be soaked up to get a non-vanishing result.
A $\frac{1}{r}$-BPS state breaks $4 \cN(1 - \frac{1}{r})$ supercharges resulting in the same number of zero modes.
Hence, the first helicity supertrace to which a multiplet of $\frac1r$-BPS states in
a 4d theory with $\cN$ extended supersymmetry can contribute non-trivially is $B_{2\kk}$ with
\begin{equation}\label{n/N}
\kk=\frac{\cN}{r}(r-1).
\end{equation}
Hence, the index $\Omi{2|2}(\gamma)$ counting $\hf$-BPS states in $\cN=2$ theories is encoded in $B_2$, whereas
the indices $\Omi{4|2}(\gamma)$ and $\Omi{4|4}(\gamma)$ in $\cN=4$ can be obtained from $B_4$ and $B_6$, respectively.
In order to extract an index from a helicity supertrace, one should factor out the center of mass contribution
equal to the helicity supertrace of the corresponding BPS multiplet,\footnote{For $\hf$-BPS states in $\cN=2$ theory
this corresponds to the contribution of a half-hypermultiplet, which is consistent with the fact that the contribution
of a single half-hypermultiplet to the BPS index $\Omega(\gamma)$ is equal to one \cite{Manschot:2010qz}.}
\be
\Bcm_{2\kk}(j)\equiv B_{2\kk}(\cR_{j,2\kk})=(-1)^{2j+\kk} 2^{-2\kk}(2\kk)!(2j+1),
\label{contr-cm}
\ee
where  $\cR_{j,2\kk}$  denotes the supersymmetry multiplet constructed by
acting on a spin $j$ ground state with $2\kk$ oscillators.
Hence we can write down indices
in $\mathcal{N}=2,4$ theories as
\be
\Omi{2|2}(\gamma)=\frac{B_2(\cH^{\cN=2}_{\gamma,j})}{\Bcm_2(j)}\, ,
\qquad
\Omi{4|2}(\gamma)=\frac{B_4(\cH^{\cN=4}_{\gamma,j})}{\Bcm_4(j)}\, ,
\qquad
\Omi{4|4}(\gamma)=\frac{B_6(\cH^{\cN=4}_{\gamma,j})}{\Bcm_6(j)}\, ,
\label{relOmHS}
\ee
where $\cH_{\gamma,j}=\cH'_\gamma \otimes \cR_{j,2\kk}$.  Here
we have suppressed the suffix $2\kk$ on $\cH_{\gamma,j}$ as it is purely determined by the charge $\gamma$
and the total number of supersymmetries, and instead indicated explicitly the latter.

At this juncture, we make the following significant observation. All of the different
helicity supertraces $B_{2\kk}$ can be obtained from a single helicity generating function \cite{Kiritsis:1997gu}
\be
B(\cR,y)=\Tr_{\cR} (-y)^{2J_3}
\ee
as
\be
B_{2\kk}(\cR)=\(\haf\, y\p_y\)^{2\kk} B(\cR,y)|_{y=1}.
\label{relBB}
\ee
One immediately recognizes this generating function to be reminiscent of the refined
BPS index $\Omega(\gamma,y)$ introduced in \eqref{defOmr}.
Indeed, to obtain the latter, one needs only to factor out again the center of mass contribution
\be
\Omega(\gamma,y)=\frac{B(\cH_{\gamma,j},y)}{\Bcm(j,y)}\, .
\label{relOmrShg}
\ee
Since, by definition, in the unrefined limit, the refined index reduces to
the $\hf$-BPS index of $\cN=2$ theory, $\Omega(\gamma,1)=\Omi{2|2}(\gamma)$, the center of mass contribution must
be given by the helicity generating function evaluated on an $\cN=2$ short multiplet which results in
\be
\Bcm(j,y)\equiv B(\cR_{j,2},y)=(-1)^{2j+1}\,\frac{y^{2j+1}-y^{-2j-1}}{y-y^{-1}}\, \frac{(y-1)^2}{y}\, .
\label{Bcmr}
\ee
It is easy to check that, setting $y=1$ in \eqref{relOmrShg}, ones does reproduce
the first relation in \eqref{relOmHS} because in $\cN=2$ theory
the helicity generating function behaves near $y=1$ as $B(\cH^{\cN=2}_{\gamma,j},y)\sim (y-1)^2$.

Importantly, the relation \eqref{relOmrShg} holds as a perfectly sensible relation beyond $\cN=2$.
Thus, the refined index can be defined in a theory with any number of supersymmetries through
its relation to the helicity generating function and the only information about
its $\cN=2$ origin is contained in its normalization factor \eqref{Bcmr}.
It can also be related to the indices counting $\frac1r$-BPS multiplets in
theories with $\cN$ extended supersymmetry with generic $\cN$ and $r$, which we define by generalizing \eqref{relOmHS}
\be
\Omi{\cN|r}(\gamma)=\frac{B_{2\kk}(\cH^{\cN}_{\gamma,j})}{\Bcm_{2\kk}(j)}\, ,
\label{relOm2nHS}
\ee
where $\kk$ is determined by $\cN$ and $r$ by \eqref{n/N}.
To get such a relation, one should note that $B(\cR_{j,2\kk},y)\sim (y-1)^{2\kk}$ near $y=1$.
This fact together with the relation \eqref{relBB} ensures that $B_{2\kk}$ is the first non-vanishing helicity supertrace
for such multiplets.
By combining \eqref{relBB}-\eqref{Bcmr}, one can then
express the index \eqref{relOm2nHS} through the refined BPS index as
\be
\Omi{\cN|r}(\gamma)=\frac{(-1)^{\kk-1} }{(2\kk-2)!}\, \p_y^{2\kk-2} \Omega(\gamma,y)|_{y=1}.
\label{relOmN}
\ee
In particular, for $(\cN|r)=(4|2)$ and $(4|4)$ corresponding to $\kk=2$ and $3$ respectively,
this provides a precise relation between the BPS indices in $\cN=4$ theory and the refined
BPS index introduced in the $\cN=2$ context. More generally, eq. \eqref{relOmN} encodes the formula
for extracting BPS indices in a general theory with extended
supersymmetry from a single refined index.

\subsection{Proposal}

Given that one can write down a refined BPS index in any theory with extended supersymmetry, the next natural step is
to generalize the results about modularity of its generating functions, presented in section \ref{subsec-D420},
to string compactifications with $\cN>2$.
In particular, we assume that in all these theories one can still construct the generating functions $\hr_{p,\mu}$ \eqref{defhDTr}
of (rational) refined indices evaluated at the large volume attractor point and that they behave under modular transformations
as vector valued (higher depth) mock Jacobi forms.
To write however a generalization of the holomorphic anomaly equation satisfied by their completions requires
understanding of several issues.

The first new feature appearing for $\cN>2$ is that the charge lattice is larger than the homoology lattice of
the CY and includes NS charges appearing due to the existence of non-trivial one-cycles.
The full charge vector can be represented as $\gamma=(p^I,q_I)=(p^0,p^A,q_A,q_0)$ where $A$ runs over $b_2+2b_1$ values and
as before we will restrict to the vanishing D6-brane charge $p^0=0$.

The crucial role in the construction of section \ref{subsec-D420} was played by the quadratic form $\kappa_{ab}=\kappa_{abc}p^c$
determined by the intersection numbers. In theories with $\cN>2$ there is a natural generalization of this object which
can be read off from the classical prepotential governing the couplings of vector multiplets in the effective action
at the two-derivative level.
The prepotential has a cubic form\footnote{The prepotential can also have contributions quadratic in $X^I$, but
they can be removed by a symplectic transformation at the expense of making charges rational \cite{Alexandrov:2010ca}.
This is the origin of the rational shifts in \eqref{fractionalshiftsD5}.\label{foot-prep}}
\be
\Fcl(X)=-\frac{\kappa_{ABC}X^A X^B X^C}{6X^0}\, ,
\label{Fcl}
\ee
and hence defines a tensor\footnote{This tensor is invariant under U-duality group in five-dimensions
and determines the classical entropy of 5d black holes \cite{Cvetic:1996zq}.}
$\kappa_{ABC}$, which can be seen as an extension of the intersection numbers.
It defines the natural quadratic form $\kappa_{AB}=\kappa_{ABC}p^C$
which appears in the spectral flow transformations \eqref{flow} \cite{Kraus:2006nb}.
Hence, one may expect it to replace $\kappa_{ab}$ also in the definition of the quadratic form $Q_n$ \eqref{defQlr}
entering the holomorphic anomaly equation for refined indices.

This replacement has important implications.
First, it changes the rank of the quadratic form and hence the modular weight of various theta series
appearing as building blocks in the construction of the completion. On the other hand, the weight
of $\hr_{p,\mu}$ was tuned to compensate the weight of these theta series. Thus, one can expect that now it will be given by
\be
\wr=-\hf\, \rank(\kappa_{AB}).
\label{weght-ref}
\ee
In the $\cN=2$ case the rank was always maximal due to the ampleness condition on the divisor.
It turns out that for $\cN>2$ this condition should be relaxed and replaced by a weaker one that
the divisor is {\it effective}, i.e. $p^A\ge 0$.
In particular, as we will see in the next section, for $\hf$-BPS states in $\cN=4$ the corresponding divisor is never ample.
As a result, the rank can be less than $b_2+2b_1$ and
the weight \eqref{weght-ref} starts depending on the choice of the magnetic charge vector $p^A$.

If the rank is not maximal, the quadratic form is not invertible so that
we should clarify the meaning of the expressions involving its inverse, like the one for the invariant charge \eqref{defhq0}.
It turns out that in this case the electric charge lattice is reduced: the charges associated with the degenerate directions
of the quadratic form get frozen. More precisely, if $\{\lambda_s^A\}$ is the set of eigenvectors of $\kappa_{AB}$ with zero eigenvalue, i.e.
$\kappa_{AB}\lambda_s^B=0$, then the charges must satisfy
\be
\lambda_s^A q_A=0.
\label{fixq}
\ee
In fact, it is easy to see that this condition is necessary for theta series to converge. Moreover, it was noticed to hold in the case of
non-compact CYs \cite{Alexandrov:2019rth}, and in appendix \ref{ap-attr} we show that it also follows from the attractor equations.
As a result, the relevant charge lattice which one sums over in theta series is
\be
\Lambda_p=\{ q_A\in \IZ  + \frac12 \,\kappa_{AB}p^B \ :\  \lambda_s^A q_A=0\}.
\label{Lam-p}
\ee
After the reduction to $\Lambda_p$ the quadratic form becomes non-degenerate and hence invertible.
It is convenient to introduce the embedding of its inverse into $\IR^{b_2+2b_1}$ which we denote by $\kappa^{AB}$.
It can be defined by the following conditions:
i) $\rank(\kappa^{AB})=\rank(\kappa_{AB})$, ii) $\kappa^{AC}\kappa_{CB}=\delta^A_B-\sum_{s,t} e^{st} \lambda_s^A\lambda_{t,B}$
where $e^{st}$ is the inverse of $e_{st}=\lambda_s^A\lambda_{t,A}$.
Then a proper generalization from the $\cN=2$ case involves a replacement $\kappa^{ab}q_a q_b\to \kappa^{AB}q_{A}q_{B}$
where the charges are supposed to satisfy \eqref{fixq}.

Another new feature is that the signature of the quadratic form $\kappa_{AB}$, even after the restriction to $\Lambda_p$,
is not necessarily Lorentzian, i.e. it can be $(n_+,n_-)$ with $n_+>1$.
This fact drastically affects the construction of $\whhr_{p,\mu}$ because
the naive extension of the formula from \cite{Alexandrov:2019rth} to the larger lattice with the quadratic form $\kappa_{AB}$
would lead to a divergent expression. A way to cure this problem is to change the kernels of theta series like $\cJr_n$ in \eqref{exp-derwh}.
In this paper we do not provide explicit expressions for the new kernels and simply assume that they exist.
The only additional assumption which will be used here is that, as soon as the quadratic form $Q_n$
has $n-1$ negative eigenvalues,
the kernels are given by those found in \cite{Alexandrov:2019rth}.

Summarizing, we formulate the following
\begin{conj}
\label{conj}
The generating functions $\hr_{p,\mu}$ of refined BPS indices defined in \eqref{defhDTr}
with
\be
\hq_0=q_0-\hf\, \kappa^{AB}q_{A}q_{B}\le \hat q_0^{\rm max}=\frac{1}{24}(p^3+c_{2,a}p^a),
\label{defhq0new}
\ee
transform as vector valued (higher depth) Jacobi forms of weight \eqref{weght-ref}
and index\footnote{The shift by $N$ was noticed in \cite{Alexandrov:2019rth} in the context of non-compact CY threefolds.}
$\mr(p)=-\(\frac{1}{6} p^3 + \frac{1}{12} c_{2,a}p^a\)-N$
where $p^3=\kappa_{ABC} p^A p^B p^C$ and $N=b_2+2b_1-\rank(\kappa_{AB})$.
Their modular completions $\whhr_{p,\mu}$ satisfy the holomorphic anomaly equation \eqref{exp-derwh}
where $\gama=(p^A,q_A=\mu_{A}+\hf\, \kappa_{AB}p^{B})$, $\gama_i=(p_i^A,q_{i,A})$,
the magnetic charges $p^A$, $p_i^A$ are all non-negative,
the electric charges are decomposed as
\be
q_{i,A}=\kappa_{i,AB}\(\epsilon^{B} +\hf\,p^{B}\)+\mu_{A},
\qquad
\eps^{A}\in \IZ, \quad \lambda_s^A\epsilon_A=\lambda_s^A\mu_A=0,
\quad \kappa_{i,AB}=\kappa_{ABC}p_i^C,
\label{sfdecomp-d}
\ee
and
\be
Q_n(\{\gama_i\})= \kappa^{AB}q_{A}q_{B}-\sum_{i=1}^n\kappa_i^{AB}q_{i,A}q_{i,B} \, .
\label{defQd}
\ee
Finally, the kernels $\cJr_n$ are supposed to have a zero of order $n-1$ at $y=1$ and
in the case of $Q_n$ with $n-1$ negative eigenvalues to coincide with the ones found in the $\cN=2$ case.
\end{conj}

\subsection{The unrefined limit}
\label{subsec-limit}

Let us now show how the above conjecture can be used to extract modular properties of the generating functions of
the {\it unrefined} BPS indices. The key fact is the relation \eqref{relOmN}.
Applying it to \eqref{defhDTr}, one can obtain a relation between the generating functions of refined and unrefined indices,
generalizing \eqref{defhDT} valid in the $\cN=2$ case,
\be
\hi{\cN|r}_{p,\mu}(\tau)=\sum_{\hat q_0 \leq \hat q_0^{\rm max}}
\bOmi{\cN|r}_{p,\mu}(\hat q_0)\,\q^{-\hat q_0  }
=\frac{2\I(2\pi)^{3-2\kk} }{(2\kk-2)!}\, \p_z^{2\kk-2}( z \hr_{p,\mu}(\tau,z))|_{z=0},
\label{relhhr}
\ee
where as usual $\kk$ is determined by \eqref{n/N} and $\bOmi{\cN|r}_{p,\mu}$ is a linear combination of the unrefined indices
\be
\label{defbOmN}
\bOmi{\cN|r}(\gamma) = \sum_{d|\gamma} d^{2\kk-4} \,\Omi{\cN|r}(\gamma/d)
\ee
evaluated at the large volume attractor point. This combination follows directly from \eqref{defbOm}
after applying the derivative operator.
It is worth to note that for $\cN>2$, one has $\kk\ge 2$ so that, in contrast to the $\cN=2$
case, the indices \eqref{defbOmN} are {\it not} rational, although still different from $\Omi{\cN|r}(\gamma)$
for non-primitive charges.

The relation \eqref{relhhr} allows to read off the modular weight of $\hi{\cN|r}_{p,\mu}$ from that of $\hr_{p,\mu}$.
Indeed, since $z$ is an elliptic variable, i.e. transforms as a modular form of weight $-1$, this relation implies that
the generating function $\hi{\cN|r}_{p,\mu}$ is a vector valued (mock)
modular form of weight
\be
w=2\kk-3+\wr.
\label{weight-hN}
\ee
Given \eqref{weght-ref}, this agrees with the formula \eqref{weight} given in the Introduction.

The mock nature of $\hi{\cN|r}_{p,\mu}$ is characterized by the holomorphic anomaly equation for its completion
(if it is non-trivial), which should also be obtained as the limit $y\to 1$ of \eqref{exp-derwh}.
Before taking the limit, it is useful to rewrite this equation in a slightly different form
\be
(y-y^{-1})\p_{\bar\tau}\whhr_{p,\mu}(\tau,z)=\q^{\hf\,\kappa^{AB}q_{A}q_{B}} \sum_{q_0}\sum_{n=2}^\infty
\sum_{\sum_{i=1}^n \gamma_i=\gamma}\!\!
\frac{\cJr_n(\{\gama_i\},\tau_2,y)}{(y-y^{-1})^{n-1}}
\prod_{i=1}^n \(\bOm(\gamma_i,y)\, \q^{-q_{i,0}}\),
\label{exp-derwh-Om}
\ee
where we simply substituted the definition of the generating functions on the r.h.s. and used the spectral flow invariance
of the refined BPS indices. The reason for this rewriting is that the charges $\gamma_i$ that differ only
by D0-brane charge $q_{i,0}$ may correspond to states preserving different number of supersymmetries.
If this is the case, the behavior of the corresponding indices $\bOm(\gamma_i,y)$ in the unrefined limit will be different
and hence they will contribute differently to the anomaly equation in this limit.

Now it is clear how to proceed: one should apply the derivative operator from \eqref{relhhr} to both sides
of \eqref{exp-derwh-Om} and take the unrefined limit.
To this end, it is crucial to take into account that the coefficients $(y-y^{-1})^{1-n}\cJr_n$ are
assumed to be finite in the limit reducing to $\cJ_n$ \eqref{unreflim-cJ}, while the refined BPS indices
develop a zero of certain order. More precisely, in a theory with $\cN$ extended supersymmetry
they behave as $\sim z^{\cN-2}$ if the charge corresponds to a $\hf$-BPS state and for other BPS states they vanish even faster.
Thus, to get a non-vanishing contribution, all powers of $z$ must be canceled by the derivative operator.
However, the number of derivatives satisfies
\be
2\kk-2=2\cN-2-\frac{2\cN}{r}\le n(\cN-2), \quad n\ge 2,
\ee
with equality reached (for $\cN>2$) only for $n=2$ and $r=\cN$.
This simple inequality immediately implies several important consequences:
\begin{itemize}
\item
The holomorphic anomaly equation can be non-trivial, and hence generating functions be {\it mock} modular,
only for $\frac{1}{\cN}$-BPS states.

\item
Only $\hf$-BPS states can contribute to the r.h.s. of the holomorphic anomaly equation.

\item
For $\cN>2$ only the contribution of $\hf$-BPS states with $n=2$ survives the unrefined limit.

\end{itemize}
Thus, for $\cN>2$ the only non-trivial holomorphic anomaly equation takes the following form
\be
\p_{\bar\tau}\whhi{\cN|\cN}_{p,\mu}(\tau)
=\q^{\hf\,\kappa^{AB}q_{A}q_{B}} \sum_{q_0}
\sum_{\gamma_1+\gamma_2=\gamma}\!\!
\cJ_2(\{\gama_1,\gama_2\},\tau_2)
\prod_{i=1}^2 \(\bOmi{\cN|2}(\gamma_i)\, \q^{-q_{i,0}}\).
\label{anomeqN}
\ee
All other generating functions \eqref{relhhr}
must be vector valued modular forms with respect to the full $SL(2,\IZ)$ duality group.

In the next section we verify these predictions for $\cN=4$ against the known results explained in section \ref{sec-N4}
and extract a new result concerning modularity of $\frac14$-BPS states with a non-trivial torsion.

\section{Modularity of immortal dyons from refinement}
\label{sec-relation}

Let us now apply the formalism of the previous section to the compactification of type II string theory
on the CY threefold $\CY=K3\times T^2$ which preserves $\cN=4$ supersymmetry in four dimensions.
This CY is characterized by the following data
\be
b_1=2,
\qquad
b_2=23,
\qquad
c_{2,a}p^a=24 p^\flat,
\ee
where the index $\flat$ corresponds to the divisor $\gamma_\flat=[K3]$.
Thus, the indices $A,B,\dots$ run over $b_2+2b_1=27$ values $A\in \{\flat,\alpha\}=\{\flat,1,\dots,26\}$,
and the non-vanishing components of the symmetric tensor $\kappa_{ABC}$ are given by
\be
\kappa_{\flat \alpha\beta}=\eta_{\alpha\beta}
=
\(\begin{array}{cc}
I_{1,1}^{\oplus 5} & 0
\\
0 & -C_{16}
\end{array}\).
\ee
The most general charge vector which we consider takes the following form in the heterotic frame (c.f. \eqref{chD420})
\be
\gamma=\(\begin{array}{ccc}
q_0, & -p^\flat, & q_\alpha
\\
q_\flat, & 0, & \eta_{\alpha\beta}p^\beta
\end{array}\).
\label{ch-het}
\ee
For this charge, the T-duality invariants \eqref{invar} are found to be
\be
n=\hf\,q^2-p^\flat q_0 ,
\qquad
m= \hf\, p^2,
\qquad
\ell=p^\alpha q_\alpha-p^\flat q_\flat,
\label{Tinvgen}
\ee
where $q^2=\eta^{\alpha\beta}q_\alpha q_\beta$ and $p^2=\eta_{\alpha\beta}p^\alpha p^\beta$,
and one has $p^3=6mp^\flat $.

To proceed with the analysis, we need to further specify the charge vector so that it describes either
a half- or a quarter-BPS state.

\subsection{Half-BPS states}

As was recalled in section \ref{sec-N4}, $\hf$-BPS states are distinguished by the condition that the vectors given by
the first and second lines of the charge \eqref{ch-het} are parallel.
There are two possibilities to satisfy this condition.

First, if $p^\flat>0$, then all charges in the second line must vanish.
To simplify the analysis, we restrict ourselves to the case $p^\flat=1$, which ensures the primitivity of the charge
and will be enough for our purposes. Thus, we consider
\be
\gamma_1=\(\begin{array}{ccc}
q_0, & -1, & q_\alpha
\\
0, & 0, & 0
\end{array}\).
\label{BPS12-1}
\ee
In this case the quadratic form $\kappa_{AB}=\kappa_{ABC}p^C$ is
\be
\kappa_{AB}=\(\begin{array}{cc}
0 & 0
\\
0 & \eta_{\alpha\beta}
\end{array}\).
\ee
It is degenerate and has rank equal to 26.
Note that the existence of a degenerate direction in the charge lattice implies the vanishing of $q_\flat$,
consistently with the $\hf$-BPS condition. The non-degenerate part of the quadratic form is $\eta_{\alpha\beta}$
and it is unimodular.
This fact implies that there is no the residual flux $\mu$, and
the generating function $\hi{4|2}_{p_0}(\tau)$ of $\hf$-BPS states, where $p_0=(1,0,\dots, 0)$ is the magnetic charge,
is a modular scalar. Comparing \eqref{defhq0new} with \eqref{Tinvgen}, one concludes that
\be
\hq_0=-n\le \hq_0^{\rm max}=1.
\ee
Thus, the generating function can be explicitly written as
\be
\hi{4|2}_{p_0}(\tau)=\sum_{n=-1}^\infty \Omi{4|2}(n)\, \q^n
\label{countOm12}
\ee
and coincides with \eqref{countd}.
Furthermore, the formula \eqref{weight-hN} and the analysis of the holomorphic anomaly done in the previous section imply that
it is a modular form of weight $w=2\cdot 2-3 -26/2=-12$, with a trivial multiplier system as can be checked from \eqref{STref}
specialized to $p^\alpha=\mu=0$.
Since the space of cusp modular forms (i.e. those which have only positive Fourier coefficients)
of weight 12 is one-dimensional and generated by $\eta(\tau)^{24}$ \cite{Dabholkar:2012nd},
it follows that $\hi{4|2}_{p_0}(\tau)\sim \eta(\tau)^{-24}$, consistently with \eqref{Z12eta}.
Of course, the precise proportionality coefficient cannot be fixed from the analysis based only on modularity.

The second possibility to get a $\hf$-BPS state is to take $p^\flat=0$ and other components in the
two lines proportional to each other, i.e.
\be
\gamma_2=\(\begin{array}{ccc}
\frac{\epsilon}{d_Q}\, q_\flat, & 0, & \frac{\epsilon}{d_Q}\,\eta_{\alpha\beta}p^\beta
\\
q_\flat, & 0, & \eta_{\alpha\beta}p^\beta
\end{array}\),
\qquad \epsilon\in \IZ, \quad d_Q=\gcd(q_\flat, \{p^\alpha\}).
\label{BPS12-2}
\ee
In this case the charge $q_0$ is not independent, but fixed in terms of other charges.
This implies that it is impossible to construct a generating function as in \eqref{relhhr}.
Nevertheless, as we now demonstrate, the formalism of the previous section works even in  such trivial situation.

For the charge \eqref{BPS12-2}, the quadratic form reads as
\be
\kappa_{AB}=\(\begin{array}{cc}
0 & \eta_{\alpha\beta}p^\beta
\\
\eta_{\alpha\beta}p^\beta & 0
\end{array}\).
\ee
It has rank equal to 2, which corresponds to 2 integers, $q_\flat$ and $\epsilon$, labelling independent electric charges in \eqref{BPS12-2}.
Therefore, according to \eqref{weight-hN}, the generating function corresponding to the magnetic charge
$(0,p^\alpha)$ must be a modular form of weight $w=2\cdot 2-3 -2/2=0$, which implies that the generating
function is a constant!

On the other hand, what the fixation of $q_0$ really means is that, due to the $\hf$-BPS condition,
in the generating function \eqref{relhhr} only one term survives.
To find this term, note that the matrix $\kappa^{AB}$ defined below \eqref{Lam-p} is given by
\be
\kappa^{AB}=\frac{1}{p^2}\(\begin{array}{cc}
0 & p^\alpha
\\
p^\alpha & 0
\end{array}\).
\ee
Substituting it into \eqref{defhq0new} together with the constraints on $q_0$ and $q_\alpha$ implied by \eqref{BPS12-2},
it is easy to show that $\hq_0$ vanishes. Thus, consistently with the above conclusion based on modularity,
the generating function reduces to a constant term.

\subsection{Quarter-BPS states: holomorphic anomaly}

The indices counting $\frac14$-BPS states in $\cN=4$ theory are moduli dependent due to the formation and decay of bound states
across lines of marginal stability in the moduli space.
However, being evaluated at the attractor point, they coincide with the indices counting only single-centered black holes
 because the only difference might arise due to scaling solutions, but being composed
of at least three constituents they do not contribute to BPS indices in $\cN=4$ theory \cite{Dabholkar:2009dq}.
Therefore, it is natural to expect that
the indices $\Omi{4|4}_{p,\mu}(\hq_0)$ coincide with the ones captured by the function $\psi^F_m$,
which appears in the decomposition \eqref{decomppsi} and also counts the immortal dyons.
As was reviewed in section \ref{subsec-mockimmort}, $\psi^F_m$ is a mock Jacobi form with the completion $\whpsi^F_m$ satisfying
the holomorphic anomaly equation \eqref{holanom}. This nicely fits with the result of the previous section that
the generating function $\hi{4|4}_{p,\mu}$ is also mock modular. Below we demonstrate that
the holomorphic anomaly equation \eqref{anomeqN} satisfied by its completion can be used to derive the equation \eqref{holanom}
for $\whpsi^F_m$ as well as its generalization for $I(\gamma)>1$.

To achieve this goal, let us make all ingredients of \eqref{anomeqN} explicit.
We restrict our consideration to the charges \eqref{ch-het}
with $p^\flat=1$ and $p^2>0$. The first condition, as in the previous subsection, guaranties the primitivity of the charge,
while the second ensures the existence of a solution to the attractor equations.
For such charge the quadratic form is non-degenerate and is given with its inverse by
\be
\kappa_{AB}=\(\begin{array}{cc}
0 & \eta_{\alpha\beta}p^\beta
\\
\eta_{\alpha\beta}p^\beta & \eta_{\alpha\beta}
\end{array}\),
\qquad
\kappa^{AB}=\frac{1}{p^2}\(\begin{array}{cc}
-1 & p^\alpha
\\
p^\alpha & p^2\eta^{\alpha\beta}-p^\alpha p^\beta
\end{array}\).
\label{qf14}
\ee
This implies the following spectral flow decomposition of the electric charges
\be
\begin{split}
q_\flat=&\, \eta_{\alpha\beta}\epsilon^\alpha p^\beta+\mu_\flat+m,
\\
q_\alpha=&\, (\epsilon^\flat+1) \eta_{\alpha\beta} p^\beta+\eta_{\alpha\beta}\epsilon^\beta +\mu_\alpha,
\end{split}
\ee
Since $|\det \kappa_{AB}|=p^2=2m$, the residual flux $\mu$ takes $2m$ non-equivalent values which can be represented
by $\mu_\flat=0,\dots, 2m-1$ and vanishing $\mu_\alpha$.
Accepting this choice and substituting the decomposition into \eqref{Tinvgen}, one finds
\be
\ell=m-\mu_\flat+ 2m\eps^\flat
\label{resellgen}
\ee
so that this T-duality invariant changes under spectral flow by $2m$.

According to \eqref{anomeqN}, the non-vanishing contributions to the anomaly equation arise only from splits of the charge
$\gamma=\gamma_1+\gamma_2$ where $\gamma_1$ and $\gamma_2$ are both $\hf$-BPS charges.
It is easy to see that this is possible only if one of them belongs to the class \eqref{BPS12-1} and the other to \eqref{BPS12-2}.
Thus, the charges are restricted to satisfy
\be
\(\begin{array}{ccc}
q_{1,0}, & -1, & q_{1,\alpha}
\\
0, & 0, & 0
\end{array}\)
+
\(\begin{array}{ccc}
\frac{\epsilon}{I(\gamma)}\, q_{2,\flat}, & 0, & \frac{\epsilon}{I(\gamma)}\,\eta_{\alpha\beta}p^\beta
\\
q_{2,\flat}, & 0, & \eta_{\alpha\beta}p^\beta
\end{array}\)
=\(\begin{array}{ccc}
q_0, & -1, & \eta_{\alpha\beta}p^\beta
\\
\mu_\flat+m, & 0, & \eta_{\alpha\beta}p^\beta
\end{array}\).
\label{sumgam}
\ee
Here we took into account that $d_Q$ defined in \eqref{BPS12-2} coincides with the torsion invariant
$I(\gamma)$ \eqref{defI} of the full charge, which due to \eqref{resellgen} can also be written as
\be
I(\gamma)=\gcd(\ell,\{p^\alpha\}).
\label{Iell}
\ee
The constraint \eqref{sumgam} fixes all charges of the constituents in terms of the full charge and one parameter $\epsilon$.
This means that the lattice one sums over on the r.h.s. of the anomaly equation is one-dimensional!
It is immediate to check that the quadratic form \eqref{defQd} depends on the free parameter as
$Q_2=-\frac{2m}{I(\gamma)^2}\, \epsilon^2+O(\epsilon)$ and thus has signature $(0,1)$.
As a result, according to our assumption, the function $\cJ_2$ can be replaced by the simple
exponential function \eqref{cJ2} found in \cite{Alexandrov:2019rth}.
Taking into account that
\be
(pp_1p_2)=2m,
\qquad
\gamma_{12}=\ell-\frac{2m\epsilon}{I(\gamma)},
\qquad
\kappa^{AB}q_A q_B=2m-\frac{\ell^2 }{2m},
\ee
where we used \eqref{resellgen} with $\epsilon^\flat=0$, and parametrizing the residue flux as
$\mu(\ell)=(m-\ell, 0,\dots,0)$,
the anomaly equation \eqref{anomeqN} takes the following explicit form\footnote{Note the factor of 2 coming from the symmetry
$\gamma_1\leftrightarrow\gamma_2$.}
\be
\tau_2^{3/2}\p_{\bar\tau}\whhi{4|4}_{p,\mu(\ell)}(\tau)
= \frac{(-1)^\ell\sqrt{m}}{8\pi\I}
\sum_{\epsilon\in\IZ}\(\sum_{\hq_{1,0}\le 1}\bOmi{4|2}(\gamma_1)\,\q^{-\hq_{1,0}}\)
\bOmi{4|2}(\gamma_2)\, \brq^{\frac{1}{4m}\(\frac{2m\epsilon}{I(\gamma)}-\ell\)^2}.
\label{anomeqNJ2}
\ee
As follows from \eqref{countOm12}, the sum over $\hq_{1,0}$ gives the factor $\eta(\tau)^{-24}$.
To deal with the sum over $\epsilon$, we decompose it into two sums by representing
$\epsilon=s-I(\gamma) r$ where $r\in \IZ$ and $s=0,\dots, I(\gamma)-1$.
Then we note that by a duality transformation the charge $\gamma_2$
can be brought to the form
\be
\gamma'_2=\(\begin{array}{ccc}
0, & 0, & 0
\\
\frac{d_s}{I(\gamma)}\,(2m-\ell), & 0, & \frac{d_s}{I(\gamma)}\,\eta_{\alpha\beta}p^\beta
\end{array}\),
\qquad
d_s=\gcd(s,I(\gamma))\, .
\label{BPS12-2p}
\ee
In this form the charge has only one non-vanishing T-duality invariant $m'=\frac{d_s^2}{I(\gamma)^2}\, m$.
As a result, the invariance under U-duality and the relation \eqref{defbOmN} allow to express the BPS index $\bOmi{4|2}(\gamma_2)$
in terms of the coefficients of \eqref{countOm12}
\be
\bOmi{4|2}(\gamma_2)=\sum_{d|d_s}\Omi{4|2}(m'/d^2).
\ee
Finally, the sum over $r$ can be evaluated explicitly giving rise to the theta function \eqref{deftheta} at $z=0$.
Thus, the holomorphic anomaly equation becomes
\be
\tau_2^{3/2}\p_{\bar\tau}\whhi{4|4}_{p,\mu(\ell)}(\tau)
= \frac{(-1)^\ell\sqrt{m}}{8\pi\I\eta(\tau)^{24}}
\sum_{s=0}^{I(\gamma)-1}\overline{\theta_{m,\ell-\frac{2ms}{I(\gamma)}}(\tau,0)}
\sum_{d|d_s} \Omi{4|2}\(\frac{d_s^2}{d^2}\, \frac{m}{I(\gamma)^2}\).
\label{anomeq-h}
\ee

To bring the resulting equation to the same form as \eqref{holanom}, one needs to construct a Jacobi form
out of the modular vector. Comparing the modular transformations of $\whhi{4|4}_{p,\mu(\ell)}$
and the theta series $\theta_{m,\ell}(\tau, z)$,
one observes that their multiplier systems, \eqref{STour} and \eqref{eq:thetatransforms}, cancel
each other up to a factor produced by $(-1)^\ell$.
In other words, the following function transforms as a mock Jacobi form
\be
\psi_{p}(\tau,z)= \sum_{\ell=-m}^{m-1}(-1)^\ell \, \hi{4|4}_{p,\mu(\ell)}(\tau)\,\theta_{m,\ell}(\tau,z)
=\sum_{n\ge -1}\sum_{\ell\in\IZ}\Omi{4|4}_{p,\mu(\ell)}\(-\tfrac{\Delta}{4m}\) \q^n \, (-y)^\ell,
\label{defpsiI}
\ee
where we expressed the invariant charge in terms of duality invariants
\be
\hq_0=-n+\frac{\ell^2}{4m}=-\frac{\Delta}{4m}
\label{defhq0K3gen}
\ee
and took into account that $\hq_0^{\rm max}=\frac{m}{4}+1$.
Given that the modular weight of $\hi{4|4}_{p,\mu}$, as follows from \eqref{weight-hN}, is equal to
$w=2\cdot 3-3 -27/2=-\frac{21}{2}$, the function \eqref{defpsiI} has weight $-10$ and index $m$, which perfectly agrees
with the properties of the function $\psi^F_m$ defined in section \ref{subsec-mockimmort}.
Thus, we expect that in the case $I(\gamma)=1$ these two functions can be identified.
Furthermore, setting $I(\gamma)=1$ in \eqref{anomeq-h} and rewriting it in terms of $\psi_{p}$,
it is immediate to see that one recovers the holomorphic anomaly equation \eqref{holanom}.

It is important to note that, as can be seen from \eqref{Iell},
the torsion invariant depends on $\ell$, the variable one sums over in the definition \eqref{defpsiI}.
Therefore, setting $I(\gamma)=1$, we assume that this constraint holds for all $\ell=-m,\dots,m-1$.
This is equivalent to the simpler condition that $d_p\equiv \gcd\{p^\alpha\}=1$.
Once this condition holds, the anomaly equation \eqref{anomeq-h} reduces to the well known equation \eqref{holanom}.

On the other hand, if we do not impose the condition $d_p=1$ in \eqref{anomeq-h},
it provides a generalization of the anomaly equation which includes contributions of charges with a non-trivial torsion.
In the rest of this section we derive a simplified form of the resulting anomaly equation and show its consistency with
the formula \eqref{Om14-I}.

To this end, we rewrite \eqref{anomeq-h} in terms of $\psi_{p}$ and perform some manipulations with the finite sums.
First, we note that $\dd'=I(\gamma)d/d_s$ divides both $d_p$ and $\ell$, while $I(\gamma)/\dd'=d_s/d$ divides $s$.
This allows to trade the sum over $d$ for the sum over $d'$ and exchange the order of summation so that this sum appears the first.
By changing the variables as $\ell=\dd' k$ and $s=\frac{I(\ell)}{\dd'}\, c$, one obtains
\bea
\tau_2^{3/2}\p_{\btau}\,\whpsi_p(\tau,z)&=&
\frac{\sqrt{m}}{8\pi \I \eta(\tau)^{24}}
\sum_{\dd'|d_p}  \Omi{4|2}\(\frac{m}{\dd'^2}\)
\sum_{k=0}^{\frac{2m}{\dd'}-1}\theta_{m,\dd' k}(\tau,z)
\sum_{c=0}^{\dd'-1}\overline{\theta_{m,\dd' k-\frac{2mc}{\dd'}}(\tau,0)}.
\label{holan-dgen}
\eea
Next, we manipulate with the sum of theta series as follows (we drop the prime on $\dd$ to avoid cluttering)
\be
\begin{split}
&\, \sum_{k=0}^{\frac{2m}{\dd}-1}\theta_{m,\dd k}(\tau,z)
\sum_{c=0}^{\dd-1}\overline{\theta_{m,\dd k-\frac{2mc}{\dd}}(\tau,0)}
\\
=&\,\sum_{k=0}^{\frac{2m}{\dd}-1}\sum_{r\in 2m\IZ+\dd k}\q^{\frac{r^2}{4m}}\, y^r
\sum_{c=0}^{\dd-1}\sum_{r'\in 2m\(\IZ-\frac{c}{\dd}\)+\dd k}\brq^{\frac{r'^2}{4m}}
\\
=&\, \sum_{\ell=0}^{\frac{2m}{\dd^2}-1}\sum_{s=0}^{\dd-1}\,
\sum_{r\in 2m\(\IZ+\frac{s}{\dd}\)+\dd \ell}\q^{\frac{r^2}{4m}}\, y^r
\sum_{r'\in \frac{2m}{\dd}\,\IZ+\dd \ell}\brq^{\frac{r'^2}{4m}}
\\
=&\,\sum_{\ell=0}^{\frac{2m}{\dd^2}-1}
\overline{\theta_{\frac{m}{\dd^2}, \ell}(\tau,0)}\sum_{r\in \frac{2m}{\dd}\,\IZ+\dd \ell}\q^{\frac{r^2}{4m}}\, y^r
= \sum_{\ell=0}^{\frac{2m}{\dd^2}-1}\overline{\theta_{\frac{m}{\dd^2}, \ell}(\tau,0)}\,
\theta_{\frac{m}{\dd^2},\ell}(\tau,\dd z)
\end{split}
\label{sumths}
\ee
where at the second step we decomposed $k=\ell+\frac{2ms}{\dd^2}$.
As a result, the anomaly \eqref{holan-dgen} for $d_p>1$ can be expressed
through the anomaly for $d_p=1$: using the function $\cA_{m}(\tau,z)$ defined in \eqref{holanom} as the r.h.s.
of the holomorphic anomaly equation for trivial torsion, the general anomaly equation takes the form
\be
\tau_2^{3/2}\p_{\btau}\,\whpsi_p(\tau,z)
= \sum_{\dd|d_p} \dd \,\cA_{m/\dd^2}(\tau,\dd z).
\label{mainanomaly}
\ee

Let us now check the consistency of this anomaly with the formula \eqref{Om14-I} which
expresses the degeneracies for $I>1$ through those for $I=1$.
Using the identification
\be
\Omi{4|4}_{p,\mu(\ell)}\(-\tfrac{\Delta}{4m}\)=\Omi{4|4}_{I,\star}(n,m,\ell),
\ee
where $I$ is given by \eqref{Iell}, and substituting \eqref{Om14-I}, the mock Jacobi form \eqref{defpsiI}
can be rewritten as
\be
\begin{split}
\psi_{p}(\tau,z)=&\,
\sum_{n\ge -1}\sum_{\ell\in \IZ}\sum_{\dd | (\ell,d_p)} \dd \, \Omi{4|4}_{1,\star}\(n,\frac{m}{\dd^2},\frac{\ell}{\dd}\)\, \q^n \,(-y)^\ell
\\
=&\, \sum_{\dd|d_p} \dd\sum_{n\ge -1}\sum_{\ell\in \dd\IZ}\Omi{4|4}_{1,\star}\(n,\frac{m}{\dd^2},\frac{\ell}{\dd}\) \q^n \,(-y)^\ell
\\
=&\,
\sum_{\dd|d_p} \dd\,\psi_{m/\dd^2}^F(\tau,\dd z).
\end{split}
\label{phipsi-gen}
\ee
It is clear that this result implies the holomorphic anomaly equation for $\whpsi_{p}$
which coincides with the one in \eqref{mainanomaly} derived from the formalism based on the refinement.

It is worth emphasizing that the function $\psi_p$ for $d_p>1$ comprises contributions of charges with {\it different}
values of the torsion invariant: the sum over $\ell$ allows us to get all $I$ dividing $d_p$.
In this sense $\psi_p$ is very different from the function obtained by expanding the generating function $\Zi{4|4}_I$
with a fixed torsion. This difference is the reason why the former is a mock Jacobi form
for the {\it full} $SL(2,\IZ)$ group, whereas the latter transforms properly only under the congruence subgroup $\Gamma^0(I)$.
This is a generic feature of our construction in that it automatically produces functions
which transform as (mock) modular or Jacobi forms under the full $SL(2,\IZ)$.
It tells us how charges with different torsion should be combined together in order to form objects with nice modular properties.
In particular, we expect that generalizing our results for $p^\flat>1$, even more sophisticated combinations of BPS indices can be
turned into $SL(2,\IZ)$ modular functions.

\section{Discussion}
\label{sec-concl}

In this paper we established a relation between the results encoding modular properties of the generating functions
of D4-D2-D0 black holes in $\cN=2$ string compactifications, and their refined version,
with the corresponding results in compactifications with $\cN>2$ extended supersymmetry.
Key to this relation is the observation that the refined BPS index
$\Omega(\gamma,y)$ coincides with a properly normalized
helicity generating function, so that all BPS indices $\Omi{\cN|r}(\gamma)$
counting $\frac{1}{r}$-BPS states in a theory with $\cN$ supersymmetries
can be obtained from $\Omega(\gamma,y)$ by taking an appropriate number
of derivatives with respect to the refinement parameter $y$
and consequently setting it to $y=1$ (see \eqref{relOmN}).
This allows to extract information about the modular properties of generating functions
of $\Omi{\cN|r}(\gamma)$ from the corresponding generating functions of $\Omega(\gamma,y)$
by providing a direct link between them, and also between the holomorphic anomaly equations satisfied by their completions.
In particular, assuming the holomorphic anomaly equation for $\whhr_{p,\mu}(\tau,z)$
to have a universal form independent of the number $\cN$ of extended supersymmetries,
one finds easily that in an $\cN >2$ theory the unrefined limit
of the holomorphic anomaly can be non-trivial only for $r=\cN$, in which case it
receives contributions only from two $\hf$-BPS states.

Next, we tested the predictions of this refined construction approach against known results
on counting BPS states in $\cN=4$ string theory on $K3\times T^2$,
and found perfect agreement for both half- and quarter-BPS states.
More precisely, the generating function for half-BPS states was shown to be uniquely determined by
the modular properties predicted for it by the refined counting, whereas for quarter-BPS states
this approach accurately reproduced the holomorphic anomaly equation of \cite{Dabholkar:2012nd}
satisfied by the completion of the generating function of immortal dyons.
Finally, we also derived a generalization \eqref{mainanomaly} of this equation which comprises contributions
of BPS dyons with non-trivial torsion $I(\gamma)>1$ and demonstrated its consistency with the formula of \cite{Dabholkar:2008zy}
expressing $\Omi{4|4}_I$ through the degeneracies for $I=1$.

In order to appreciate the difference between BPS state counting in $\cN = 2$ and $\cN >2$ theories,
it useful to recall that for the latter there exists a description of BPS degeneracies
via generating functions that transform as modular objects under the U-duality group of the theory.
Furthermore, mock modularity arises when one restricts oneself to counting only a specific class of BPS states
such as single centred or immortal black holes.
There is no analogous description in the $\cN=2$ case.  The generating functions
obtained in our approach are not invariant under the full U-duality group, as can be seen from the fact that
they require vanishing of the D6-brane charge. Instead, they preserve only a subgroup of
the T-duality group and transform generically as mock modular forms under $SL(2,\IZ)$.

This is where our approach comes into its own. Due to its universality,
it naturally produces generating functions which turn out to be modular or mock modular
with respect to the full $SL(2,\IZ)$ group, irrespective of the number $\cN$ of supersymmetries.
In contrast, the naive extension of the counting of $\frac14$-BPS states to non-trivial torsion $I>1$,
based on duality invariant generating functions,
leads to modular forms under a smaller congruence group $\Gamma^0(I)$ \cite{Dabholkar:2008zy}.
The modularity under $SL(2,\IZ)$ is achieved due to a clever combination of contributions with different torsion.
It is the formalism explained above that ensures the right combination, which otherwise would
have been difficult to arrive at by plain guesswork.

There are natural avenues for extending the results reported here.
Firstly, in the $\cN=4$ case we imposed a restriction on the charge counting D4-branes wrapped on $K3$, $p^\flat\le 1$.
Although this is enough to reproduce the known counting functions, it would be interesting to relax this restriction.
One may expect that the generating functions with $p^\flat > 1$ provide new examples of non-trivial
modular combinations of BPS indices,
which might even lead to new interesting mathematical constructions of (vector valued) modular forms.

Secondly, this approach should be tested in other cases where there are known results on counting BPS states.
One such possibility is to consider the CHL orbifolds \cite{Chaudhuri:1995fk,Dabholkar:2006bj}
(see \cite{Bossard:2018rlt,Fischbach:2020bji,Cardoso:2020swg} for the current state of the art of these models).
Another option is given by compactification
of type II string theory on $T^6$ leading to $\cN=8$ supersymmetry, where
a plethora of rigorous results (see e.g. \cite{Shih:2005qf,Pioline:2005vi,Sen:2008ta,Sen:2008sp}) furnishes
a rich test-ground for the predictions of our approach.

Finally, we have left an important gap in our construction. Namely, due to the non-standard signature of the quadratic form $\kappa_{AB}$
appearing in compactifications with $\cN>2$ extended supersymmetry, we did not determine the explicit form of the functions $\cJr_n$
entering the holomorphic anomaly equation in the refined case.
This explicit form is not required to get the unrefined limit for $\cN=4$, and it is likely that the same is still true for $\cN=8$.
However, this gap should be filled in order, in particular, to confirm a few mild assumptions we made about $\cJr_n$.
This problem also represents an interesting mathematical challenge with the promise of generating
new insights into the theory of indefinite theta series, to which string theory has already made important contributions.

\section*{Acknowledgements}

The authors are grateful to Gabriel Cardoso, Finn Larsen, Sameer Murthy and Boris Pioline
for valuable discussions and correspondence. SN is supported by UIDB/04459/2020.

\appendix

\section{Dualities and charges}
\label{ap-dual}

In this appendix we describe the interpretation of the charge vector \eqref{charge-het}, defined originally in the heterotic frame,
in other frames which can be related to the heterotic one by a chain of duality transformations.
Denoting $n$-dimensional cycles of the compactifications manifold by $X_n$, we will use the following abbreviations:
\\ \indent
KK/$X_1$ --- momentum state along $X_1$;
\\ \indent
F1/$X_1$ --- fundamental string wrapped on $X_1$;
\\ \indent
D$p$/$X_p$ --- D$p$-brane wrapped on $X_p$;
\\ \indent
M2/$X_2$ --- M2-brane wrapped on $X_2$;
\\ \indent
M5/$X_5$ --- M5-brane wrapped on $X_5$;
\\ \indent
NS5/$X_5$ --- NS5-brane wrapped on $X_5$;
\\ \indent
KKM$\to X_1$ --- Kaluza-Klein monopole localized in $X_1$;
\\ \indent
$(E_9\oplus E_8)_{e/m}$ --- electric or magnetic charges with respect to the gauge fields of the ten-dimensional heterotic theory.

In the heterotic frame, the compactification manifold is $T^6$ which we decompose as $T^3\times \tT^3$.
We denote one-dimensional cycles lying inside $T^3$ and $\tT^3$ by $S_r^1$ and $\tS_r^1$ with $r=1,2,3$, respectively,
and five-dimensional cycles by $T^5_r$ and $\tT^5_r$ such that $T^5_r\times S_r=\tT^5_r\times \tS_r^1=T^6$.
Then the charge components have the following interpretation
\be
\gamma=
\(\mbox{\small\begin{tabular}{@{}c@{,\,}c@{,\,}c@{,\,}c@{,\,}c@{,\,}c@{,\,}c@{}}
KK/$S_1^1$ & 1F/$S_1^1$ & $\cdots$ & KK/$\tS_1^1$ & 1F/$\tS_1^1$ &
$\cdots$ &  $(E_8 \oplus E_8)_e$
\\
NS5/$T^5_1$ & KKM$\to S_1^1$ & $\cdots$ & NS5/$\tT^5_1$ & KKM$\to \tT^5_1$ & $\cdots$ &
$(E_8 \oplus E_8)_m$
\end{tabular}}\),
\label{chahet}
\ee
where dots denote similar entries with $r=2,3$.
The same theory can be viewed as M-theory compactified on $K3\times T^3$.
From this point of view, the charge components are now interpreted as
\be
\gamma=
\(\mbox{\small\begin{tabular}{@{}c@{,\,}c@{,\,}c@{,\,}c@{}}
KK/$S_1^1$ & M5/$K3\times S_1^1$ & $\cdots$ &  $M2/\gamma^i$
\\
M2/$S^1_2\times S^1_3$ & KKM$\to S_1^1$ & $\cdots$ &
$M5/\gamma^i\times T^3 $
\end{tabular}}\),
\label{chaM}
\ee
where $\gamma^i$, $i=1,\dots,22$, are 2-cycles of $K3$.
Let us now do reduction along $S^1_1$ to the type IIA string theory on $K3\times T^2$. Then the charge becomes
\be
\gamma=
\(\mbox{\small\begin{tabular}{@{}c@{,\,}c@{,\,}c@{,\,}c@{,\,}c@{,\,}c@{,\,}c@{}}
D0 & D4/$K3$ & KK/$S_2^1$ & NS5/$K3\times S_2^1$ & KK/$S_3^1$ & NS5/$K3\times S_3^1$ &  D2/$\gamma^i$
\\
D2/$T^2$ & D6/$K3\times T^2$ & F1/$S^1_3$ & KKM$\to S_2^1$ & F1/$S^1_2$ & KKM$\to S_3^1$ &
D4/$\gamma^i\times T^2 $
\end{tabular}}\).
\label{chIIA}
\ee
This is precisely the charge vector \eqref{chD420} where it is written in terms of notations
introduced in the beginning of section \ref{subsec-charge}.
The minus sign in front of $p^\flat$ originates in the symplectic transformation
relating the type IIA and the heterotic frames \cite{Ceresole:1995jg}.

It is instructive also to establish connection with D5-D1-KK system widely used in \cite{Dabholkar:2012nd}
and other studies of string theory on $K3\times T^2$. To this end, let us return to the M-theory picture represented by the charge \eqref{chaM}
and reduce it instead along $S^1_3$. This gives another type IIA formulation compactified on $K3\times \tT^2$ with $\tT^2=S^1_1\times S^1_2$
and the corresponding charge is
\be
\gamma=
\(\mbox{\small\begin{tabular}{@{}c@{,\,}c@{,\,}c@{,\,}c@{,\,}c@{,\,}c@{,\,}c@{}}
KK/$S_1^1$ & NS5/$K3\times S_1^1$ & KK/$S_2^1$ & NS5/$K3\times S_2^1$ & D0 & D4/$K3$ & D2/$\gamma^i$
\\
F1/$S^1_2$ & KKM$\to S_1^1$ & F1/$S^1_1$ & KKM$\to S_2^1$ & D2/$\tT^2$ & D6/$K3\times \tT^2$ &
D4/$\gamma^i\times \tT^2 $
\end{tabular}}\).
\label{chIIA2}
\ee
Finally, we apply T-duality along $S_2^1$ which maps to the type IIB frame
and leads to the following interpretation of the charge vector
\be
\gamma=
\(\mbox{\small\begin{tabular}{@{}c@{,\,}c@{,\,}c@{,\,}c@{,\,}c@{,\,}c@{,\,}c@{}}
KK/$S_1^1$ & KKM$\to S^1_2$ & F1/$S_2^1$ & NS5/$K3\times S_2^1$ & D1/$S^1_2$ & D5/$K3\times S^1_2$ & D3/$\gamma^i\times S^1_2$
\\
KK/$S^1_2$ & KKM$\to S_1^1$ & F1/$S^1_1$ & NS5/$K3\times S_1^1$ & D1/$S^1_1$ & D5/$K3\times S^1_1$ &
$D4/\gamma^i\times S^1_1$
\end{tabular}}\).
\label{chIIB}
\ee
Comparing \eqref{chIIB} with \eqref{chIIA}, one observes that switching on KK momentum localized in $S^1_2$
and D1, D5-brane charges associated to $S^1_1$, as in \cite{Dabholkar:2012nd},
corresponds to having non-vanishing $p^\flat$ and $p^2$ in \eqref{chD420},
which were precisely our conditions on the charge of a $\frac14$-BPS state.

It may also be useful to note that this chain of dualities maps D1 and D5 charges to NS charges in \eqref{chIIA}.
If one wishes to map them to D-brane charges, one can perform before the last step a mirror symmetry on $K3$,
which effectively reduces to T-duality on one of its 2-cycles, say $\gamma^1$.
This intermediate step would result in exchanging the 5th and 6th columns in \eqref{chIIB} with two columns
in the last entry corresponding to $\gamma^1$ and its dual cycle $\gamma^2$ in the sense that $\gamma^1\cap \gamma^2=1$.

\section{Modular transformations}
\label{ap-modular}

In this appendix we collect transformation properties of various modular objects appearing in the main text.

First of all, let us recall the definition of a vector valued Jacobi form of weight $w$ and index $m$ \cite{MR781735}.
This is a finite set of functions $\phi_\mu(\tau, z)$ with $\tau\in \IH$, $z\in \IC$ labelled by $\mu$
such that
\be
\begin{split}
\phi_{\mu}(\tau,z+k\tau+\ell)=&\, e^{- 2\pi\I m \( k^2\tau + 2 k z\)} \,\phi_{\mu}(\tau,z),
\\
\phi_{\mu}\(\frac{a\tau+b}{c\tau+d}, \frac{z}{c\tau+d}\)
=&\, (c\tau+d)^w \, e^{\frac{2\pi\I m c z^2}{c\tau+d}}\sum_\nu M_{\mu\nu}(\rho)\,\phi_{\nu}(\tau,z),
\end{split}
\label{Jacobi}
\ee
where $\rho=\scriptsize{\(\begin{array}{cc}
a & b \\ c & d
\end{array}\)}\in SL(2,\IZ)$ and we allowed for a non-trivial multiplier system $M_{\mu\nu}(\rho)$.
Similarly, the transformations of a vector valued modular form $\phi_\mu(\tau)$ can be obtained from \eqref{Jacobi} by setting $z=m=0$
and thus they are completely specified by weight $w$ and multiplier system $M_{\mu\nu}(\rho)$.
Since $M_{\mu\nu}(\rho)$ must furnish a representation of the group and $SL(2,\IZ)$ is generated by two transformations,
$T=\scriptsize{\(\begin{array}{cc}
1 & 1 \\ 0 & 1
\end{array}\)}$
and
$S=\scriptsize{\(\begin{array}{cc}
0 & -1 \\ 1 & 0
\end{array}\)}$,
to define the multiplier system, it is enough to specify it for $\rho=T$ and $S$.
Thus, below to characterize the modular behaviour of an object, we will provide $w$, $M_{\mu\nu}(T)$, $M_{\mu\nu}(S)$
and, if necessary, $m$.

\begin{itemize}
\item
The completions $\whhr_{p,\mu}(\tau,z)$ of the generating functions of refined BPS indices are vector valued Jacobi forms
of weight $\wr$, index $\mr(p)$ and the multiplier system
\be
\begin{split}
M_{\mu\nu}(T)=&\, \delta_{\mu\nu}\, e^{\pi\I\( \frac{1}{12}\, c_{2,a} p^a +(\mu+\frac12 p)^2\)},
\\
M_{\mu\nu}(S)=&\, \frac{(-\I)^{\wr+1}}{\sqrt{|\Lambda^*/\Lambda|}}\,
e^{-2\pi\I\(\frac{1}{4}\, p^3 +\frac18\, c_{2,a}p^a\)  }\,
e^{-2\pi\I \mu \cdot \nu}\,,
\end{split}
\label{STref}
\ee
where $\mu\cdot \nu=\kappa^{AB}\mu_A\nu_B$ is determined by the quadratic form of the non-degenerate part of the lattice
of electric charges, which can be defined as
$\Lambda=\{\eps^{A}\in \IZ\ :\ \lambda_s^A\epsilon_A=0\}$.
In the $\cN=2$ case, $\Lambda=H_4(\CY,\IZ)$ and the quadratic form coincides with $\kappa_{ab}$.

\item
The completions $\whhi{\cN|r}_{p,\mu}(\tau)$ of the generating functions of unrefined BPS indices are vector valued modular forms
of weight \eqref{weight-hN} and the same multiplier system as in \eqref{STref}.
Let us make it explicit for $(\cN|r)=(4|4)$ specified by the charge \eqref{ch-het} with $p^\flat=1$.
In this case, the quadratic form is given in \eqref{qf14}, $p^3=6m$, $c_{2,a}p^a=24$, $\wr=-27/2$
and $\mu=\mu(\ell)=(m-\ell, 0,\dots,0)$.
Substituting these data into \eqref{STref}, one obtains
\be
M_{\mu(k)\mu(\ell)}(T)= \delta_{k\ell}\, e^{-\frac{\pi\I}{2m}\, \ell^2},
\qquad
M_{\mu(k)\mu(\ell)}(S)= \frac{(-1)^{\ell+k}}{\sqrt{-2m\I}}
\,e^{\frac{\pi\I}{m}\, k\ell}\,.
\label{STour}
\ee

\item
The theta series $\theta_{m,\ell}(\tau, z)$ \eqref{deftheta} is a vector valued Jacobi form of weight $1/2$, index $m$
and the multiplier system
\be
M_{k\ell}(T)=\delta_{k\ell}\, e^{\frac{\pi\I}{2m}\, \ell^2},
\qquad
M_{k\ell}(S)= \frac{e^{-\frac{\pi\I}{m}\, k\ell}}{\sqrt{2m\I }}\,.
\label{eq:thetatransforms}
\ee

\end{itemize}

\section{Attractor equations and electric charges}
\label{ap-attr}

In this appendix we analyze the attractor equations in the large volume limit
and will be particularly interested in the case where the quadratic form defined by the classical part of the prepotential is degenerate.
The attractor equations are the equations of motion of the low-energy effective action restricted to
BPS black hole near-horizon backgrounds \cite{Ferrara:1995ih}.
They allow to fix the values of scalar fields at the black hole horizon in terms of charges
and can easily be formulated in terms of the prepotential $F(X)$ as
\begin{equation}
\Im X^I= p^I,
\qquad
\Im \p_{X^I}F=q_I.
\label{el}
\end{equation}

The full quantum prepotential can be formally written in a perturbative worldsheet genus expansion
with order $g$ term captured in terms of a genus-$g$ topological string amplitude
\begin{equation} \label{top}
F(X) = \sum_{g=0}^\infty g_{\rm top}^{2g-2} F_g(z),
\end{equation}
where $g_{\rm top}= 1/{X^0}$ and $z^A=X^A/X^0$.
In the large volume limit $t^A=\Im z^A \gg 1$, the leading term in the prepotential is holomorphic and
has a universal form
\begin{equation}
F_0(z)= -\frac{1}{6}\, \kappa_{ABC}z^A z^B z^C + O(e^{-z}),
\label{F0}
\end{equation}
where the exponentially suppressed contributions correspond to worldsheet instantons.
Sometimes, the prepotential also contains contributions quadratic and linear in $z^A$,
but they can be removed by a symplectic transformation (see footnote \ref{foot-prep}).
In contrast, the first subleading term $F_1$ satisfies a holomorphic anomaly equation.
Its solution can be found explicitly and in the large volume limit
it asymptotes to a linear term (see e.g. \cite{Huang:2015sta})
\begin{equation}
F_1(z) \to  a_A z^A  + O(e^{-z}),
\end{equation}
where $a_A$ are real and depend on the second Chern class of the underlying CY.
The higher $F_g$ are exponentially suppressed in $t^A$ and so are their derivatives with respect to $z^A$.
Due to this, the generic prepotential in the large volume limit reduces to
\be
F^{\rm lv}(X)=-\kappa_{ABC}\, \frac{X^A X^B X^C}{6 X^0}+ a_A \,\frac{X^A}{X_0} .
\label{aproxF}
\ee

Let us now turn to the attractor equations.
To ensure the large volume limit, we choose the electro-magnetic charge to be
\be
\gamma=\(0,\lambda p^A,\lambda q_A,\lambda^3 \hat q_0+\frac{\lambda}{2}\,\kappa^{AB}q_A q_B\),
\label{scalegam}
\ee
where we set D6-brane charge to zero, as everywhere in the paper, and take the parameter $\lambda$ to be large.
In the case of degenerate $\kappa_{AB}=\kappa_{ABC}p^C$, the inverse $\kappa^{AB}$ is defined as below \eqref{Lam-p}.
The first attractor equation in \eqref{el} is solved by
\be
X^0\in \IR,
\qquad
X^A=\phi^A+\I \lambda p^A,
\quad \phi^A\in\IR.
\ee
Substituting this solution into the second set of equations, one finds
\bea
q_A&\approx &-\frac{1}{2\lambda X^0}\, \Im (\kappa_{ABC} X^B X^C-2a_A)=-\frac{1}{X^0}\,\kappa_{AB}\phi^B,
\label{eqqA}
\\
\hat q_0&\approx &\frac{1}{6\lambda^3 (X^0)^2}\, \Im (\kappa_{ABC}X^A X^B X^C-6a_A X^A)-\frac{1}{2\lambda^2}\,\kappa^{AB}q_A q_B
\nn\\
&=& -\frac{1}{6\lambda^2 (X^0)^2}\(\lambda^2 p^3+6 a_Ap^A\).
\label{ateq-a}
\eea
As a result, the attractor equations are solved by
\be
\begin{split}
X^0 =&\,\(\frac{ p^3+6 \lambda^{-2} a_A p^A}{-6\hat q_0}\)^{1/2},
\\
X^A=&\, -X^0\kappa^{AB}q_B+\I \lambda p^A.
\end{split}
\label{solatr}
\ee
In particular, for $p^3> 0$, $X^0\sim const$ and one reproduces the large volume attractor point \eqref{zatr}
for the moduli $z^A=X^A/X^0$ up to a redefinition of the scaling parameter and replacement $\kappa^{ab}\to \kappa^{AB}$.
On the other hand, if $p^3=0$, while the solution is still non-degenerate due to taking into account the one-loop correction $a_A$,
the scaling is different leading to $X^0\sim \lambda^{-1}$. However, the solution for the moduli
$z^A$ is still consistent with \eqref{zatr} upon redefinition $\lambda^2\to \lambda$.\footnote{Alternatively,
in this case one can choose to scale $\hat q_0$ in \eqref{scalegam} as $\lambda$, which would result in $X^0\sim const$ and
$z^A$ given in \eqref{zatr}.}

Most importantly, the attractor equation \eqref{eqqA} shows that if the quadratic form
$\kappa_{AB}$ is degenerate, a solution exists only if the electric charges satisfy certain constraints.
Namely, if $\lambda^A_s$ is a set of vectors such that $\kappa_{AB}\lambda_s^B=0$, then the existence of solution requires
\be
\lambda_s^A q_A=0.
\ee
If this is the case, eq. \eqref{solatr} with $\kappa^{AB}$ defined as below \eqref{Lam-p} provides the corresponding solution.

\providecommand{\href}[2]{#2}\begingroup\raggedright\endgroup


\end{document}